# Global 3D simulation of the upper atmosphere of HD189733b and absorption in metastable HeI and Lyα lines

M. S. Rumenskikh[134], I. F. Shaikhislamov[134], M. L. Khodachenko[2], H. Lammer[2], I. B. Miroshnichenko[14], A. G. Berezutsky[13], L. Fossati[2]

[1] Institute of Laser Physics SB RAS, Novosibirsk, Russia

[2] Space Research Institute, Austrian Academy of Sciences, Graz, Austria

[3] Institute of Astronomy, Russian Academy of Sciences, Moscow, Russia

[4] Novosibirsk State Technical University, Novosibirsk, Russia

*E-mail address: marina_rumenskikh@mail.ru*

**Abstract:** A 3D fully self-consistent multi-fluid hydrodynamic aeronomy model is applied to simulate the hydrogen-helium expanding upper atmosphere of the hot Jupiter HD189733b, and related absorption in the Lyα line and the 10830 Å line of metastable helium. We studied the influence of a high-energy stellar flux, stellar wind, and Lyα cooling to reproduce the available observations. We found that to fit the width of the absorption profile in 10830 Å line the escaping upper atmosphere of planet should be close to the energy limited escape achieved with a significantly reduced Lyα cooling at the altitudes with HI density higher than $3\cdot10^6$ $cm^{-3}$. Based on the preformed simulations, we constrain the helium abundance in the upper atmosphere of HD189733b by a rather low value of He/H~0.005. We show that under conditions of a moderate stellar wind similar to that of the Sun the absorption of Lyα line takes place mostly within the Roche lobe due to thermal broadening at a level of about 7%. At an order of magnitude stronger wind, a significant absorption of about 15% at high blue shifted velocities of up to 100 km/s is generated in the bowshock region, due to Doppler broadening. These blue shifted velocities are still lower than those (~200 km/s) detected in one of the observations. We explain the differences between performed observations, though not in all the details, by the stellar activity and the related fluctuations of the ionizing radiation (in case of 10830 Å line), and stellar wind (in case of Lyα line).

## 1. Introduction

Gas giants are numerous and most studied among the known extrasolar planets, because they produce strong easily detectable transit signals. These so-called ‘hot Jupiters’ are exposed to an intense X-ray and ultraviolet (XUV) radiation coming from the host stars and the planetary upper atmospheres are subject to strong radiative heating and ionization. This results in expansion and hydrodynamic escape of the upper atmosphere and thus mass loss (Lammer et al. 2003). Furthermore, the fact of transit gives one a possibility of detecting absorption by the upper atmosphere, probing size, composition, temperature and velocity of the surrounding media.

The first exoplanet atmosphere detected with transit method was HD209458b (Charbonneau et al. 2000). Its transits were also observed in Lyα line (Vidal-Madjar et al. 2003) and revealed absorption several times higher than that in optic photometry (Ben-Jaffel 2007; Vidal-Madjar et al. 2008). Later, Vidal-Madjar et al. (2004) Linsky et al. (2010) and Cubillos et al. (2020) measured the comparable transit absorption depths of 6-9 % for this exoplanet in the CII, OI, SiIII and FeII lines. These observations of HD209458b were followed by similar spectral measurements for the HD189733b, which also transits a bright star. Due to the greater planet/star radii ratio, it is even more suitable for transit measurements. However, in other aspects HD189733b is quite different from HD209458b. Most notably, it is smaller, but more massive, and it orbits significantly closer to its active K1.5V host star (Bouchy 2005).

Lecavelier des Etangs (2010, 2012) reported one spectrally unresolved absorption in the Lyα line at the level of 5.05±0.75% and one resolved absorption with almost the same value of the total absorption of 5.0±1.3% over the whole line, but with an essentially enhanced absorption of 14.4±3.6% in the range of blue-shifted velocities [-230; -140] km/s. Etangs et al. (2012) pointed out that this effect might be related to the upper atmosphere of HD189733b which overflows the Roche lobe and supersonically streams away. There was also one spectrally resolved measurement, which yielded a very small excess absorption in the Lyα line (Etangs et al. 2012). Besides of that, the far ultraviolet (FUV) absorption lines of ionized C, N, Si have shown significant short-time variability, most probably caused by flare activity of the host star (Pillitteri et al.2015). This makes the analysis and interpretation of the transit absorption by this exoplanet in FUV lines particularly difficult and ambiguous.

An important feature of HD189733b consists in a severe day-night heating contrast. The simulations with a 3D global circulation model (GCM) predicted the zonal winds with velocities [-5; 10] km/s in the lower layers of the atmosphere (Showman & Polvani 2011). The upper atmosphere escape on HD189733b was modeled with 1D hydrodynamic (HD) codes. In particular, for the ionizing ($\lambda$<912 Å) XUV radiation flux of $F_{XUV}=2\cdot10^4$ erg $cm^{-2/}$s at the planet orbit, Guo (2011) obtained the planetary mass loss rate of $5\cdot10^{10}$ g/s, the upper atmosphere maximum temperature slightly above $10^4$ K, and the location of half-ionization $H/H^+=1$ level at ~$1.1R_p$, indicating a very fast photo-ionization of the atmospheric hydrogen. Menager et al. (2013), using an aeronomy code, found that the $H/H^+=1$ level appears at ~$1.5R_p$. The most detailed and recent 1D HD simulation by Salz et al. (2016) predicts for $F_{XUV}=1.8\cdot10^4$ erg $cm^{-2}$/s a somewhat lower mass loss rate of about $(1\text{-}2)\cdot10^{10}$ g/s, the maximum temperature of $1.2\cdot10^4$ K, and the location of $H/H^+=1$ level at $1.2R_p$. As a reason for the lower mass loss rate, was found the cooling effect, caused by the excitation of hydrogen and emission of Lyα photons.

There are only a few works dedicated to the interpretation of the Lyα observations of the HD189733b. Bourrier et al. (2013) used a 3D Monte Carlo code, tracing the hydrogen atoms launched from an inner spherical boundary, surrounding the planet at $3R_p$. They found that acceleration of the atoms by radiation pressure, alone is not sufficient for the generation of energetic neutral atoms (ENA) population with the velocities of up to 200 km/s, and an additional mechanism should operate, namely charge exchange between the atmospheric atomic hydrogen and stellar protons. In Odert et al. (2020) a 1D HD model of planetary wind was combined with a 3D MHD model of the magnetized stellar wind (SW). For the XUV flux of $1.8\cdot10^4$ erg $cm^{-2}$/s, they obtained the upper atmosphere maximum temperature of $\approx1.1\cdot10^4$ K, the mass loss rate of $5\cdot10^{10}$ g/s, and the location of $H/H^+=1$ level at $\approx1.7R_p$. Odert et al. (2020) also found that the absorption in Lyα line is mostly due to the natural line broadening mechanism, and is of ~11% in the blue wing of the line [−400; −40] km/s, and 13% in the red wing [40; 400] km/s. Both numbers are higher than the observed value. Odert et al. (2020) have shown that even an extremely strong SW would not generate enough ENAs to produce large blue shifted absorption in excess of 140 km/s reported in Lecavelier des Etangs et al. (2010, 2012). The parameters of the considered strongest SW corresponded to the conditions in CMEs with the velocity of 1000 km/s and the integral stellar mass loss $10^3$ times higher than the average solar value ($\approx2.5\cdot10^{12}$ g/s). Under these conditions the magnetopause was pushed by the SW ram pressure towards the planet as close as $2.3R_p$.

Since 2018 a new opportunity for probing of the expanding and escaping planetary atmospheres was opened (Seager & Sasselov 2000; Oklopčić & Hirata 2018), based on the measurements of the transit absorption at the position of metastable helium $2^3$S triplet line at 10830 Å (hereafter HeI($2^3$S)). The metastable $2^3$S level of helium atom has a high excitation energy of 19.8 eV and a radiative life-time of about 2.2 hours. It is efficiently pumped by the recombination of ionized helium and results in the absorption in infrared band due to the triplet transition $2^3S$-$2^3P$ at 10830.34 Å, 10830.25 Å, and 10829.09 Å. The HeI($2^3$S) 10830 Å triplet line has been used to estimate the magnetic field in solar filaments (Lin et al. 1998) and to determine the helium abundance in extragalactic HII regions (Aver et al. 2015), as well as to probe the material flows in active galactic nuclei (Leighly et al. 2011). Seager & Susselov (2000) proposed using this line to observe the transits of HD209458b. Determination of the population of HeI($2^3$S) state in the upper atmospheres of hot exoplanets is a rather complex problem tackled in Oklopčić & Hirata (2018) by means of 1D HD model. They predicted the absorption at the line center at a level of 8% and 2% for GJ436b and HD209458b, respectively.

Recent IR observations of exoplanetary transits with ground telescopes, mostly with CARMENES instrument on the Calar Alto and NIRSPEC on the Keck II, provided a series of valuable data for the analysis and interpretation. The HeI($2^3$S) absorption depths of ~8% for the Wasp107b (Allart et al. 2019; Kirk et al. 2020), ~1% for the HATP11b (Allart et al. 2018, Mansfield et al. 2018), GJ3470b (Ninan et al. 2020, Pallé et al. 2020), and HD209458b (Alonso-Floriano et al. 2019), as well as ~3% for the Wasp69b (Nortmann et al. 2018), were measured. Most of the absorption profiles have a typical spectral width of ±10 km/s in Doppler velocity units. The reproducibility of the experimental data, the typical signal-to-noise ratio and the statistical significance $\geq10\sigma$ noticeably exceed those achieved for the measurements in FUV band. As to the hot Jupiter HD189733b, considered in the present paper, there are two independent observations described by Salz et al. (2018) and Guilluy et al. (2020). They are similar in half-width of the HeI($2^3$S) absorption profile, but differ in depths by a factor of 1.5 (i.e., 1 and 0.7 %, respectively). There is also a net blue-shift of the whole absorption profile by ≈2.5 km/s reported in Salz et al. (2018), but absent in Guilluy et al. (2020) data. We note that signatures of a small blue-shift is also reported for other planets. However, Salz et al. (2018) remarks that these blue-shift velocities could be potentially caused by the stellar pseudo-signals, i.e. due to the stellar surface regions with a stronger HeI absorption. To simulate the absorption profiles at 10830 Å for different exoplanets a number of 1D HD models were used. These 1D models calculate temperature, velocity, electron, hydrogen and helium density distribution profiles in the upper atmosphere and the kinetics of HeI($2^3$S) state population. Ninan et al. (2019) and Palle et al. (2020) found for the GJ3470b that the area of HeI($2^3$S) presence extends up to at least $10R_p$, and the observed transit depths can be fitted in the modeling with a relatively low helium abundance of about He/H~0.01. The most sophisticated simulation of the HeI($2^3$S) absorption, based on 1D HD modeling is described by Lampón et al. (2021). Because we explored the GJ3470b in our previous work (Shaikhislamov et al. 2020), and consider HD189733b in this paper, we discuss the approach of Lampón et al. (2021) in more detail. It is based on a Parker-like solution, where instead of temperature, the sound speed is assumed to be constant. Besides of temperature, other fitting parameters are the mass loss rate and the helium abundance. The fixed XUV flux of $5.7\cdot10^4$ erg cm$^{-2}$/s is used in the model to calculate the ionization balance and the kinetics of HeI($2^3$S) state. To fit the measured absorption profiles, Lampón et al. (2021) used up to three velocities (that is, line-of-sight components) with different coverage of stellar disk for the HD189733b (altogether, six empirical fitting parameters), whereas for the GJ3470b only the net blue shift appeared to be sufficient. However, the fitting of the HeI($2^3$S) absorption profiles was strongly dependent on the temperature and mass loss considered as the modeling free parameters, but was equally good for the broad range of helium abundances, from 1 to 10 %. To constrain this parametric degeneracy, more self-consistent modeling approaches have to be used by Lampón et al. (2021). They used temperature and the mass loss derived from the best fit of the Parker-like solution profiles to the profiles from Salz et al. (2016, 2018) for GJ3470b, and from Salz et al. (2016) and Odert at al. (2020) for HD189733b. It is argued in Lampón et al. (2021) that to derive the He/H ratio, the parameters of planetary outflow should be constrained by the Lyα measurements. At the same time, the detected Lyα absorption of GJ3470b, is significantly blue-shifted (35±7 % in the range [-94; -41] km/s, Bourrier et al. 2018) and it cannot be generated just by the planetary atmosphere outflow. Therefore, like in the case of GJ3470b, the contribution of SW and the generation of ENAs are essential. These factors, however, cannot be properly included in the 1D model. Nevertheless, to give at least a hint of them, Lampón et al. (2021) perform one of the fits of the He($2^3$S) absorption profile with the assumption that the escaping planetary material beyond the Roche lobe moves away from the star at a constant prescribed velocity.

Thus, combining different 1D models still requires, in order to address the essentially 3D processes and observational features, artificially adding of the non-spherically symmetric effects, which only a global 3D multi-component model can simulate and predict realistic density distributions of the major species and the structure of the interacting planetary and SW material flows. This is especially true for such objects as HD209458b and HD189733b, for which the spectral absorption measurements in lines of different species have been made, like e.g., in HI (Lyα), OI, CII, HeI($2^3$S). Ideally, the interpretation of these measurements should be based on the simulation of dynamical behavior of all species and respective lines within the frame of a global self-consistent model. In the present paper we make a step forward on this way and perform a multi-fluid 3D HD self-consistent modeling of the escaping upper atmosphere of HD189733b surrounded by the stellar wind, simulated with the same model, and calculate the synthetic transit absorption features.

Besides of the simulation and interpretation of observations of HD189733b, we compare our results with other modeling approaches.

The model we use in our study here has been already applied for the simulation and interpretation of observations of GJ436b (Khodachenko et al. 2019), π Men c (Shaikhislamov et al. 2020a), HD209458b (Shaikhislamov et al. 2019, Khodachenko et al. 2021b), GJ3470b (Shaikhislamov et al. 2020b), and Wasp107b (Khodachenko et al. 2021a). The last three most recent papers also modeled the absorption by metastable helium at 10830 Å. The interpretation of observations of GJ3470b in both Lyα and HeI($2^3$S) lines revealed that bowshock layer formed by SW around the planet plays significant role. They found the best fit parameters for both lines as $F_{XUV}$=8 erg cm$^{-2}$/s, He/H=0.013, and the SW with density and velocity similar to the fast solar wind, strong enough to form a bowshock at about 20$R_p$ (Shaikhislamov et al. 2021).

The simulation of absorption at 10830 Å during the transits of Wasp107b revealed another feature which is beyond the capacity of 1D models; it concerns the radiation pressure, acting on metastable HeI($2^3$S) atoms. As pointed out in Allart et al. (2019), the acceleration of helium atoms by IR (10830 Å) radiation flux of the star Wasp107 is ~75 times higher than that due to the stellar gravity. At the same time, as shown in Khodachenko et al. (2021) the observed HeI($2^3$S) absorption at Wasp107b can be well reproduced with the typical expected parameters of the system and appropriate account of the electron and atom impact processes, without invoking of any specific additional assumptions on the stellar radiation and helium abundance.

Regarding the results reported in the present paper, we note that the 3D aeronomy simulations and global HD modeling are performed for the first time in the study of HD189733b. It is also the first attempt to interpret simultaneously both Lyα and HeI($2^3$S) measurements at transits of HD189733b within a single model. In this respect it is worth to mention that Odert et al. (2020), aiming at the interpretation of Lyα observations during the transits of HD189733b, considered a more sophisticated magnetized SW case in 3D, but they took the escaping planetary atmospheric flow from the 1D simulations and did not take helium into account at all. In the present paper, we investigate as well, like in Odert et al. (2020), weather a strong SW can generate sufficient amount of ENAs to affect the Lyα blue wing absorption and to explain the significant difference of the measured Lyα absorption profiles in different observations. Based on the performed simulations and fitting to the measured HeI($2^3$S) absorption profile, we also constrain the helium abundance value in the upper atmosphere of HD189733b and show that the differences in observations can be explained by the temporal variations of the stellar XUV flux.

The paper is organized as follows. Section 2 contains short description of the model and provides information about initial parameters; Section 3 presents the obtained results, which are discussed in Section 4.

## 2. Modeling approach

The 3D multi-fluid global model used in this paper is being developed since Shaikhislamov et al. (2014). Its most important parts are described in the following papers: the details of gasdynamic equations and terms related to generation of planetary wind, collisional interaction of species and applicability of hydrodynamic approach are given in the Shaikhislamov et al. (2016) and in the appendix of Dwivedi et al. (2019); the details of 3D code and of stellar plasma wind simulation – in Khodachenko et al. (2019); procedures of calculating transit absorption in lines – in Shaikhislamov et al. (2018b); simulation of metastable helium population and absorption – in Shaikhislamov et al. (2021a). Here we only briefly outline the model. It solves numerically the system of mass, momentum and energy equations in spherical geometry, taking into account the gravity and Coriolis forces, as well as the radiation pressure of the stellar Lyα and IR flux, acting on HI and HeI($2^3$S) atoms, respectively. The model includes also the processes of hydrogen and helium photochemistry, in particular, the reactions of photoionizaton, recombination, and electron impact. For the upper atmosphere of HD189733b, we neglect the molecular species that are present in the lower atmosphere, which are quickly dissociated with height. Thus, in this paper we deal with the following species: H, H+, He, He+, He2+. An additional argument in support of the atomic upper atmosphere assumption comes from the modeling of HeI($2^3$S) absorption discussed later in the paper. However, in several dedicated model runs we studied the effects of molecules and included $H_2$, $H_2^+$, and $H_3^+$ into consideration.

The metastable HeI($2^3$S) atoms are treated in the model as a separate fluid with its own velocity and temperature, determined by those of the species from which they originate, i.e., $He^+$ in case of recombination, or HeI($1^1$S), in case of excitation from the ground state. The reactions involved in the processes of population and depopulation of the HeI($2^3$S) triplet level are similar to those considered in Oklopčić & Hirata (2018) and described in Shaikhislamov et al. (2021a).

The code uses an explicit numerical scheme of the second order spatial and temporal accuracy, realized on a spherical grid in the planet-centered reference frame with polar axis Z, perpendicular to the orbital plane. More details about the model implementation can be found in Shaikhislamov et al. (2018a, 2020a, 2021) and Khodachenko et al. (2019, 2021). At the inner boundary of the simulation domain, i.e., the planet surface at $r=R_p$, we fix zero radial velocity of the atmospheric material, the base atmosphere temperature of 1000 K and pressure of 0.05 bar. The photo-ionization and heating of gas by photo-electrons, as well as the radiation attenuation in the planetary atmosphere, is calculated over the XUV and NUV spectra binned by 1 Å according to wavelength-dependent cross-sections of various elements involved in simulation. For the particular spectra we employ the rescaled spectral energy distribution (SED) of epsilon Eri to emulate the emission of HD189733. This choice is motivated by the fact, that the two stars have the same spectral type and comparable activity levels (see, for example, Chadney et al. 2017). In the XUV band ($10<\lambda<912$ Å) the radiation flux of this spectra is $F_{XUV} \approx 25$ erg cm$^{-2}$/s at the reference distance of 1 a.u. The latest reanalysis of available X-ray and FUV observations and modeling of synthetic EUV spectra of HD189733 (Bourrier et al. 2020) produced stellar XUV fluxes varying from visit to visit from 23 to 31 erg cm$^{-2}$/s at 1 a.u. At the same time, we distinguish in our model a hard energy part of the flux (XUVH) at $\lambda<510$ Å with $F_{XUVH}=16$ erg cm$^{-2}$/s at 1 a.u., which influences the metastable helium population. The ratio of $F_{XUVH}/F_{XUV}=2/3$ was kept unchanged in course of the modeling (except of specially addressed cases), whereas the value of total integral XUV flux $F_{XUV}$ was varied for different model runs. The employed SED is characterized in the near-ultraviolet part (NUV, $912<\lambda<3000$ Å, effects HeI($2^3$S) population by its photo-ionization) by $F_{NUV}=1830$ erg cm$^{-2}$/s at 1 a.u., while in the and near-infrared part (NIR, around 10830 Å, is used to calculate the radiation pressure force acting on HeI($2^3$S)) by $F_{10830}=20$ erg cm$^{-2}$/s Å$^{-1}$ at 1 a.u.

The absorption is calculated post-simulation by a special procedure described in Khodachenko et al. (2017) and Shaikhislamov et al. (2018b, 2020a) for Lyα line and in Shaikhislamov et al. (2021a) for HeI $2^3P$-$2^3S$ transitions. For the latter, the synthetic spectral profiles are obtained, like those in observational data of Salz et al. (2018) and Guilluy et al. (2020). That is, by averaging the absorption between the second and the third contact transit points in a planet reference frame.

## 3. Simulation results

### 3.1 The role of the Lyα cooling of upper atmosphere

The cooling of expanding upper atmospheric material due to excitation of hydrogen atoms and emission of the Lyα photons is an important effect in partially ionized upper atmospheres of hot Jupiters. However, its quantitative description is non-trivial because of the trapping and diffusion of the Lyα photons in dense regions of the planetary atmosphere. The first aeronomy codes (Yelle et al. 2004, García Muñoz et al. 2007, Koskinen et al. 2007) ignored this Lyα cooling effect, assuming eventual loss of Lyα photons to photo-ionization of the excited states of trace elements.

However, Murray-Clay et al. (2009) simulated the escaping planetary wind with the account of Lyα cooling and have shown that it changes the energy budget of a typical hot Jupiter by an order of 10%. Later, Shaikhislamov et al. (2014), using a simple model of the diffusion of Lyα photons, have shown that the collisional electron de-excitation of the excited hydrogen 2p states reduces the effect of Lyα cooling and influences the temperature profile of the upper atmospheric material near the inner boundary of the simulation domain at low altitudes. For the sufficiently massive planets the increasing effect of Lyα cooling suppresses the expansion of the planetary atmosphere, until the regime of the material dynamical escape is replaced by the quasi-static radiatively cooled atmosphere (Weber et al. 2018). Salz et al. (2016) obtained for the HD189733b exactly this regime of a quasi-stationary upper atmosphere with low mass-loss rate of $(1\text{-}2)\cdot10^{10}$ g/s.

At the same time, there exist much more sophisticated simulations of the stationary upper atmosphere of HD189733b based on the direct Monte-Carlo modeling (Christie et al. 2013 and Huang et al. 2017). This approach takes into account a number of processes, which terminate the Lyα scattering cycle. Among them are photoionization from the n=2 state, collisional de-excitation, two photon decay, and photoionization of metals, e.g., Na and Mg. Altogether, such Monte-Carlo simulation led to the conclusion that the major cooling processes involve metals and other trace elements, while the contribution of Lyα cooling does not exceed 10%, as compared with the total XUV heating.

Therefore, so far there is no model, which combines all necessary physical processes needed for the detailed and self-consistent simulation of Lyα cooling of the expanding upper atmosphere of a hot Jupiter. For a reason, which will be specified later, we consider in this work the effectiveness of the Lyα cooling depending on an empirical parameter defined via so-called density cut-off in the following way. We reduce the Lyα cooling rate by a scaling factor $\exp(-\sigma H n_H/\alpha)$, where $n_H$ is atomic hydrogen density, $\sigma\approx6\cdot10^{-14}$ cm$^2$ is cross-section of Lyα photon absorption at a typical temperature of $10^4$ K, and H is barometric scale-height at a given gas temperature. The empirical parameter α defines the number of scatterings after which the Lyα cycle is terminated. Thus, the typical value $n_{cut}$=α/σH corresponds to the density layer below which the Lyα cooling rapidly decreases in the model.

We found out that in the case of Lyα cooling acting over all altitudes the hydrodynamic escape in our model gets very low and the upper atmosphere becomes close to the radiative hydrostatic equilibrium. In this regime, the total rates of Lyα-emitted and XUV-absorbed energy are related as $W_{Ly\alpha}/W_{XUV}$=0.76 with $W_{XUV}\approx 4\cdot 10^{24}$ erg/s. The corresponding mass loss rate is $\approx 1.5\cdot 10^{10}$ g/s, maximum temperature $\approx 10^4$ K, and velocity at 2$R_p$ about 1.5 km/s. All these values are rather close to those obtained in Salz et al. (2016), despite the differences between the 1D and 3D models. However, without account of the Lyα cooling the mass loss rate becomes an order of magnitude higher, $2.2\cdot 10^{11}$ g/s, with the temperature maximum $1.5\cdot 10^4$ K, and velocity at 2 $R_p$ about 6 km/s. The differences between these two cases are shown in Figure 1. Altogether, the performed comparison reveals that without Lyα cooling the mass loss rate and velocity of material escape are higher, mostly because of the higher temperature.

The most important point here is, that the regimes with and without Lyα cooling are very much different with respect to the HeI($2^3$S) absorption, which prompted us to reconsider the problem of Lyα cooling. Under the conditions of full Lyα cooling, the velocity of escaping material (PW) is low, and the width of the corresponding HeI($2^3$S) absorption profile appears to be too small to fit the observations, whereas without Lyα cooling, the absorption width agrees with the measurements. Figure 2 (left panel) shows the simulated absorption profiles at the position around 10830 Å for different Lyα cooling (including full- and no- cooling cases), obtained with different values of the defined above cut-off density, which controls the effective trapping of Lyα photons. It shows that to make the shape of the synthetic HeI($2^3$S) absorption profiles closer to the measurements, we should assume significant reduction of the Lyα cooling, with the cut-off density at least $n_{cut}=3\cdot 10^6$ cm$^{-3}$, which corresponds to the trapping parameter $\sigma H n_H > 10^2$. In this case, the energy rates are related as $W_{Ly\alpha}/W_{XUV}$ =0.37. According to our approach, the case of practically no Lyα cooling corresponds to $n_{cut}=3\cdot 10^5$cm$^{-3}$, whereas full Lyα cooling, to $n_{cut}=10^{11}$cm$^{-3}$.

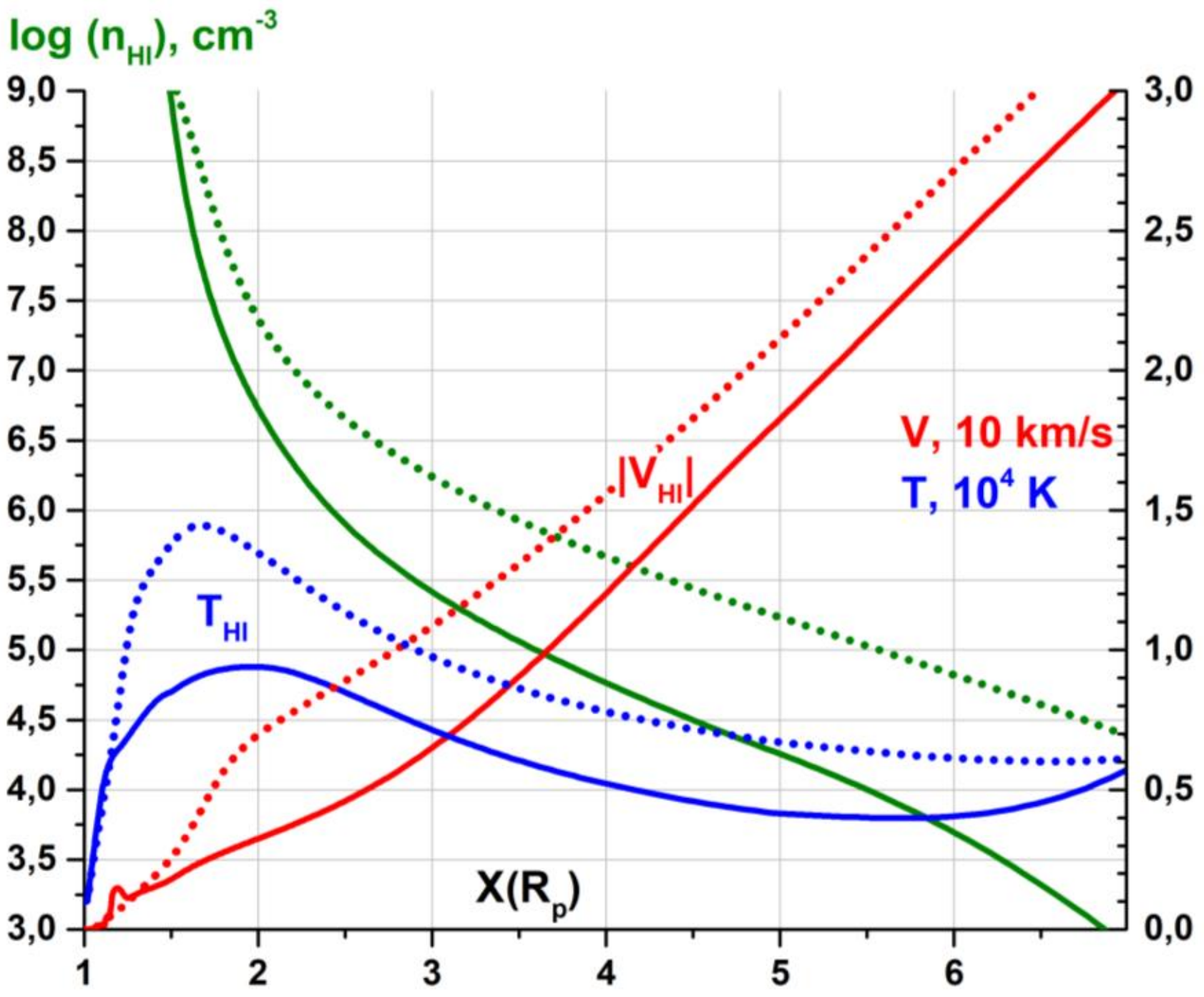

Figure 1. Density profiles of hydrogen atoms (left axis), velocity and temperature (right axis) along the planet-star line, modeled for the case of full Lyα cooling (solid lines) and no Lyα cooling (dotted lines). The distances in this plot, and in all others, are scaled in units of planet radius $R_p$.

It should be noted that our model does not show the increased absorption at the weakest component of triplet between [-50; -20] km/s, as well as the small net blue shift observed by Salz et al. (2018) (though absent in the observation of Guilluy et al. 2020). These features we discuss later. With a series of dedicated simulation runs we also checked another possibility that can produce broadening of the HeI($2^3S$) absorption profile, namely the zonal wind, whose speed for HD189733b might reach up to 10 km/s at altitudes with pressure of ≤30 mbar (Showman & Polvani 2011). For that we emulated the equatorial jet as a prescribed additional rotation of the planet by imposing the corresponding azimuthal velocity $V_\varphi = R_p \cdot \Omega \cdot \sin\theta$ at the inner boundary of the simulation domain at $r=R_p$. Here Ω is the cyclic frequency of the prescribed rotation. Right panel in Figure 2 shows the simulated absorption profiles obtained in the case of full Lyα cooling ($n_{cut} = 3 \cdot 10^{16}$ cm$^{-3}$) for three different values of the azimuthal velocity $V_\varphi$ ={19; 11; 5.5} km/s at the equator, which correspond to the additional rotation periods P=2π/Ω= {8; 12; 24} hours. In particular, it can be seen that the close to observations width of the absorption profile is achieved at rather high equatorial velocity in the range $V_\varphi$ =(10-20) km/s.

Based on this preliminary study, further on we consider only the scenario of the reduced Lyα cooling with the cut-off density $n_{cut} = 3 \cdot 10^6$ cm$^{-3}$. This value allows obtaining the most plausible widths of spectral profiles, which fit the observations at HeI($2^3S$) line.

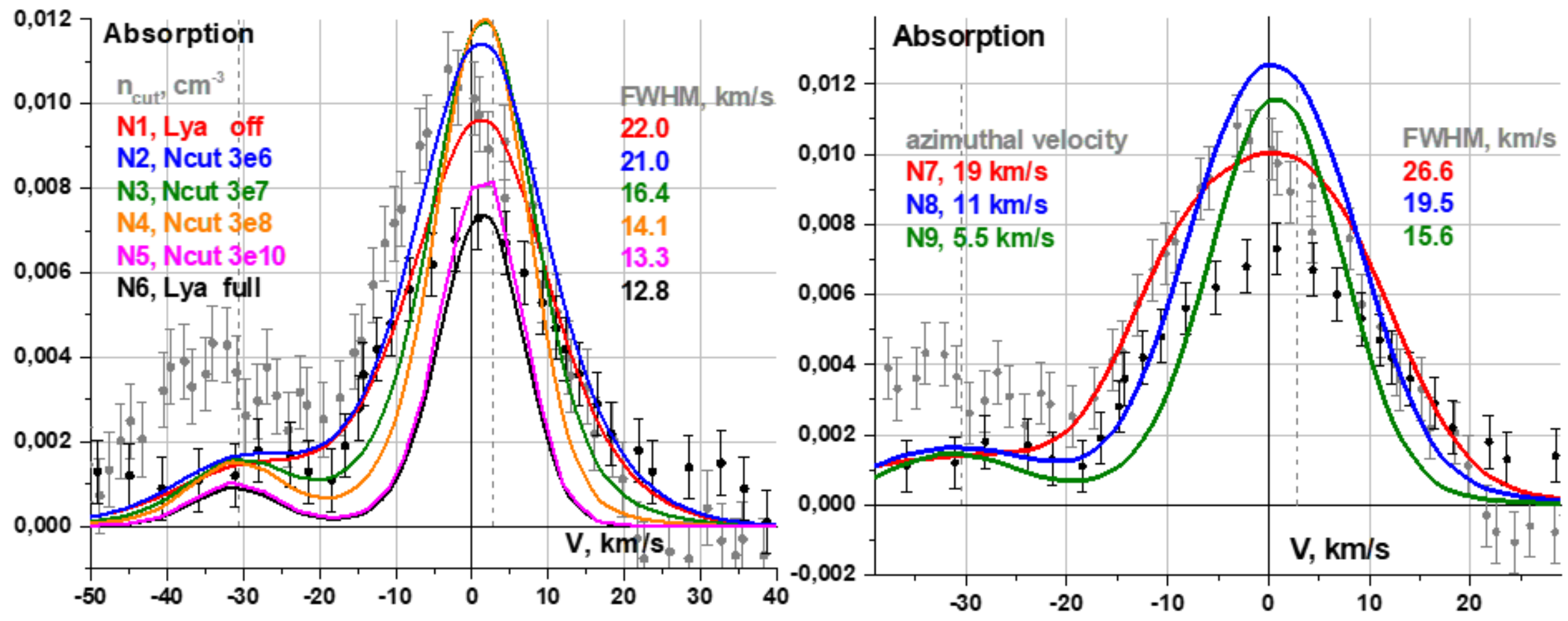

Figure 2. Left panel: HeI($2^3$S) triplet absorption profiles simulated for different degrees of Lyα cooling, expressed in terms of cut-off hydrogen density $n_{cut}$. Here and further on in similar plots the filled circles with error bars reproduce the measurements from Salz et al. (2018) (gray) and Guilluy et al. (2020) (black); vertical gray dashed lines indicate the positions of individual lines in the triplet, one of which is at V=0. Absorption in this and all other figures is dimensionless and varies in the range [0; 1]. The colored legend specifies particular simulation runs (according to Table 1 given below), the corresponding cut-off densities and widths of the absorption profile.

Right panel: HeI($2^3$S) triplet absorption profiles simulated in the case of full Lyα cooling for different planet rotation velocities (provided in legend) emulating the effect of global circulation and corresponding to the diurnal periods of 8, 12 and 24 hours.

### 3.2. Constraining XUV flux and helium abundance

At first, we consider the case of a weak SW with the following parameters at the planetary orbit: $V_{sw}$ =210 km/s, $T_{sw}$ =0.7 MK, $n_{sw}$ =$10^3$ $cm^{-3}$. It is worth to note that at the orbital distance of HD189733b, the SW is still in the process of acceleration, and orbital speed of ≈150 km/s provides a significant component to the total velocity of the planet relative the background SW plasma. Figure 3 shows the global view of the escaping upper atmosphere of the HD189733b interacting with the flow of SW plasma simulated in the model run N11 under the typical conditions specified above (see also Table 1).

The flow of atmospheric material occupies a significant area along and around the orbit, far ahead and well behind the planet. The realized in this case regime of interaction between the escaping atmospheric flow of HD189733b and the SW corresponds to the “captured by the star” regime, according to the classification introduced in Shaikhislamov et al. (2016). Because of the strong tidal and non-inertial forces, which turn the planetary material stream clockwise and align it with the orbital plane, the flow is relatively thin across the planet-star direction. Due to the stellar gravity, the stream is confined within the range ±8$R_p$ across the orbital plane.

Figure 3 (right panel) demonstrates how the HeI($2^3$S) absorption is distributed over the stellar disk as seen by the observer, i.e., the line-of-sight absorption. It takes place relatively close to the planet within a sphere of radius ~3$R_p$.

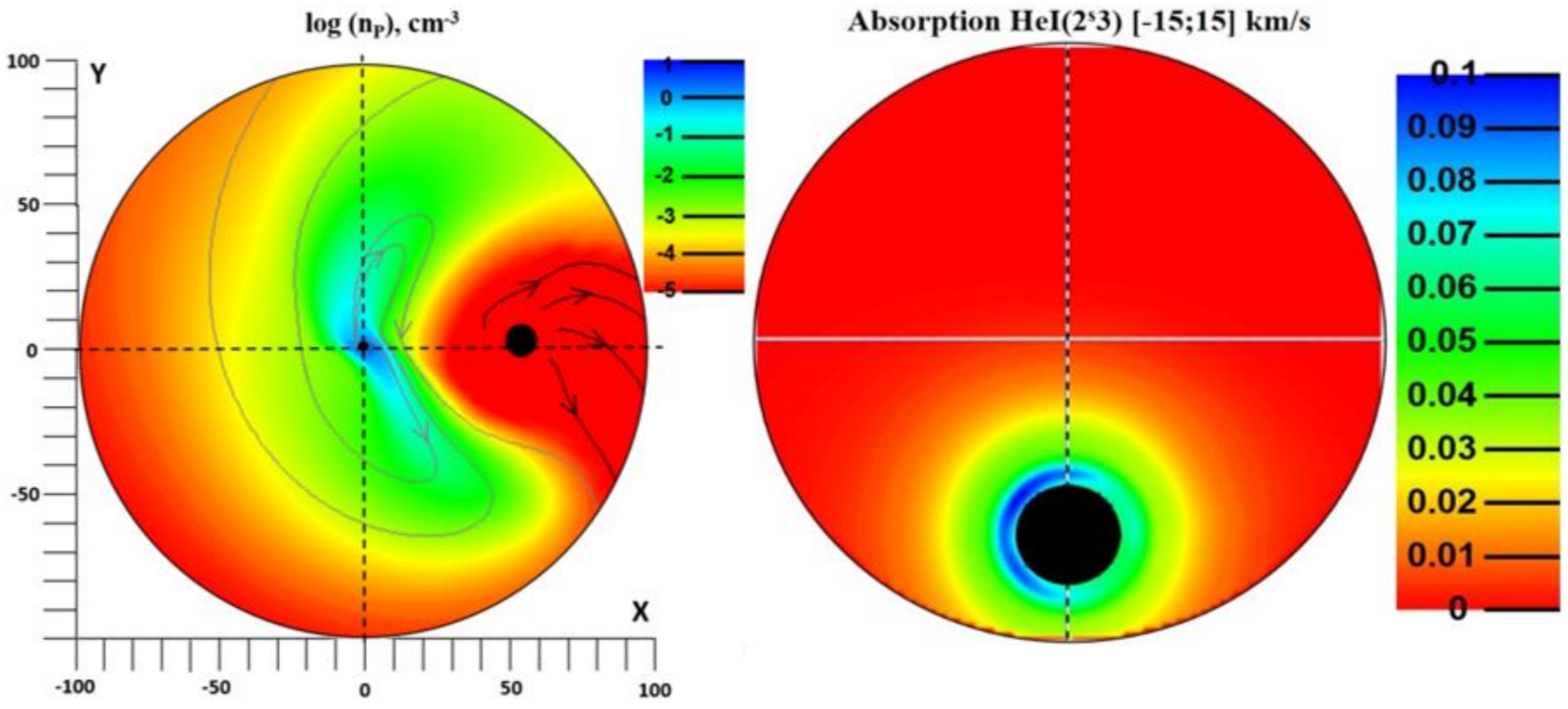

Figure 3. Left panel: Distribution of proton density around HD189733b in the orbital plane of the whole simulated domain, obtained in the simulation run N11 (see Table 1), which corresponds to the case of weak SW, no zonal wind, and reduced Lyα cooling. The planet is located at the center of coordinates and rotates anti-clockwise around the star marked as black circle at X=55. Proton fluid streamlines originated from star are shown with black lines, whereas gray lines indicate the streamlines of planetary material flow.

Right panel: The distribution of the line-of-sight absorption at the position of HeI($2^3$S) 10830 Å line integrated over ±15 km/s, as seen by a remote observer at the mid-transit. The stellar limb is the picture circumference; the planet is marked by black circle.

Left panel in Figure 4 shows detailed information on the distribution of densities of species, velocity and temperature of planetary material along the planet-star line. Due to the high gravity, the outflow velocity is relatively low in comparison to other well studied exoplanets HD209458b and GJ436b. The temperature reaches 15000 K rather close to the planet and is higher than that obtained in the 1D simulations (for example, Salz et al. 2016), which is due to the reduced Lyα cooling adopted in our modeling. Half-ionization levels for H and He are both below $2R_p$. The $H^+/H=1$ level appears at $r=1.6R_p$, which is close to that predicted in aeronomy simulations in Menager et al. (2013), but perceptibly farther than in 1D models of Guo (2011) and Salz et al. (2016). With the same value of XUV flux, we obtained for the HD189733b the mass loss rate of about three times of that reported in Guo (2011) and Odert et al. (2020).

From Figure 4 we can also surmise that PW is sufficiently collisional. For example, at a distance of $5R_p$ the helium atoms have the Knudsen number of about 0.1, calculated by collisions with protons with a cross section of about $4 \cdot 10^{-16}$ cm$^2$. Right panel shows the rates of chemical processes increasing and decreasing the metastable HeI($2^3$S) population. One can see that the recombination of ionized helium HeII into HeI($2^3$S) state is counteracted by the auto-ionization collisions with H atoms at low altitudes $r<1.3R_p$, and by the electron collisional depopulation above this height. Note that photo-ionization of HeI($2^3$S) state becomes important relatively far from the planet where the density of HeI($2^3$S) atoms becomes too low to affect the absorption. Thus, the ionizing stellar flux in the NUV band is not crucial for the HeI($2^3$S) absorption at 10830 Å for the HD189733b. In the course of modeling, by varying flux at around 10830 Å and comparing velocities, we found out that the radiation pressure causes no significant acceleration of HeI($2^3$S) atoms for this exoplanet.

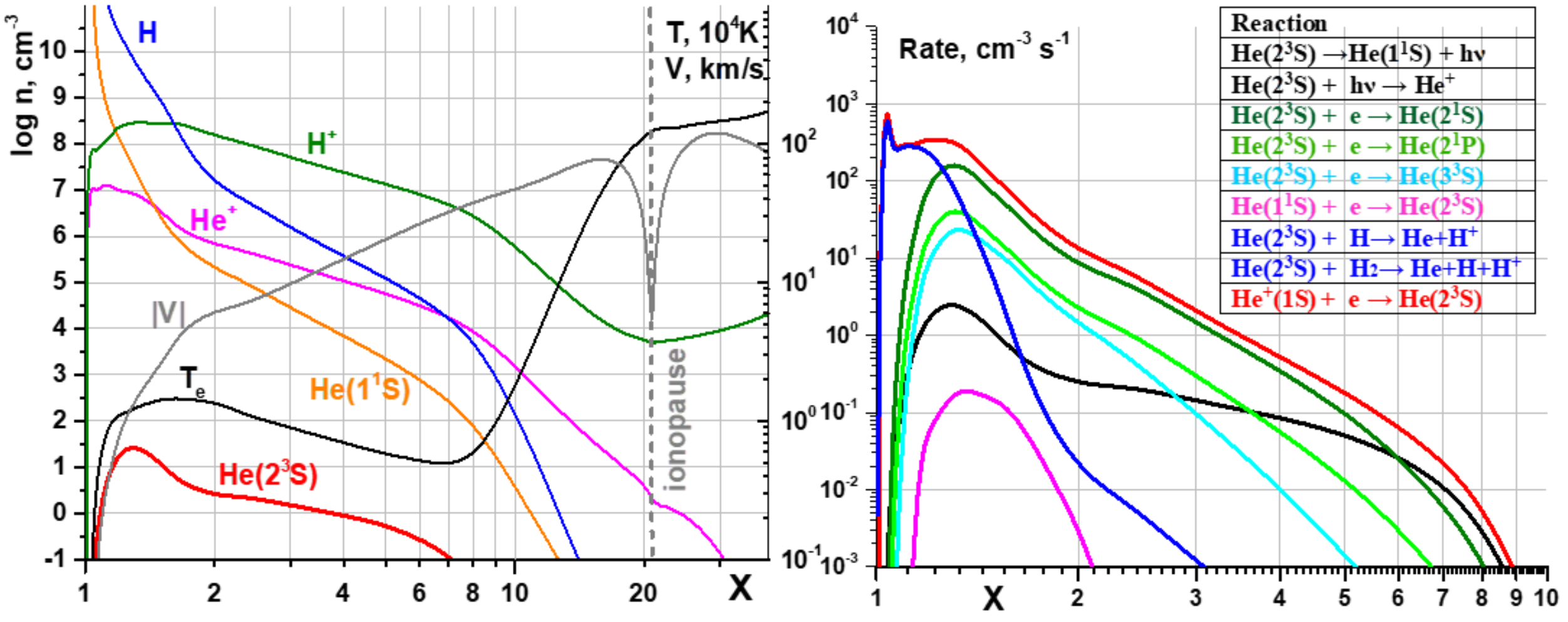

Figure 4. Left panel: Profiles of density of the major species (left axis), temperature and velocity of protons (right axis) along the planet-star line, obtained in the simulation run N11.

Right panel: Rates of the reactions responsible for the processes of population and depopulation of the metastable HeI($2^3$S). The reactions indicated in the legend with the same color are summed in the corresponding plot lines.

To find the model run which provides a best fit of observations in HeI($2^3$S) and Lyα spectral lines, we vary the value of integral stellar XUV flux, which is a subject to temporal fluctuations (Bourrier et al. 2020), and the helium abundance, which is a constant value to be constrained. Initially, we assume a very weak SW, which doesn't affect the expansion of the escaping planetary atmospheric material. Then we will compare the obtained results with the cases of more intense SW characterized by stellar mass loss rate ($M'_{sw}$).

A list of the model parameters and the corresponding absorption values calculated in different simulation runs is given in Table 1. We note that the available for HD189733b two independent measurements at HeI($2^3$S) line have significantly higher S/N ratio and similarity between each other, than the Lyα observations. Therefore, by comparison of the simulated HeI($2^3$S) absorption profiles with the measurements, we constrain at first the helium abundance. The presence and amount of helium are important for the atmospheric escape of the relatively massive HD189733b, because they influence the hydrostatic scale-height of the planetary atmosphere.

| N | $F_{XUV}$ | He/H | $M'_{sw}$ $10^{10}$ g/s | $n_{cut}$ $cm^{-3}$ | $M'_{pw}$ $10^{10}$ g/s | Abs. HeI max, % | Half-width HeI, km/s | Abs. Lyα, blue wing [-230;-140] km/s | Abs. Lyα, red wing [60;110] km/s | Abs. Lyα, total | others |
|---|---|---|---|---|---|---|---|---|---|---|---|
| 1 | 15 | 0.006 | 10 | $3.3\times10^{5}$ | 17.3 | 0.95 | 22 | 5.8 | 7.6 | 6.3 | |
| 2 | 15 | 0.006 | 10 | $3.3\times10^{6}$ | 17.1 | 1.1 | 21 | 5.9 | 8.0 | 6.8 | |
| 3 | 15 | 0.006 | 10 | $3.3\times10^{7}$ | 9.8 | 1.2 | 16.5 | 6.3 | 8.5 | 7.1 | |
| 4 | 15 | 0.006 | 10 | $3.3\times10^{8}$ | 5.8 | 1.2 | 14 | 6.4 | 7.8 | 6.9 | |
| 5 | 15 | 0.006 | 10 | $3.3\times10^{10}$ | 1.9 | 0.8 | 13.2 | 4.9 | 5.7 | 5.1 | |
| 6 | 15 | 0.006 | 10 | $3.3\times10^{16}$ | 1.6 | 0.73 | 12.8 | 4.7 | 5.4 | 4.9 | |
| 7 | 15 | 0.006 | 10 | $3.3\times10^{16}$ | 9.0 | 1.0 | 26.6 | 6.8 | 8.4 | 7.5 | $V_{\varphi}$=19 km/s |
| 8 | 15 | 0.006 | 10 | $3.3\times10^{16}$ | 8.0 | 1.25 | 19.5 | 3.9 | 4.6 | 7.7 | $V_{\varphi}$=11 km/s |
| 9 | 15 | 0.006 | 10 | $3.3\times10^{16}$ | 4.7 | 1.15 | 15.6 | 6.2 | 7.4 | 6.6 | $V_{\varphi}$=5.5 km/s |
| 10 | 7.5 | 0.006 | 10 | $3.3\times10^{6}$ | 10.6 | 0.9 | 19.1 | 6.0 | 8.3 | 6.9 | |
| 11 | 25 | 0.006 | 10 | $3.3\times10^{6}$ | 24.7 | 1.35 | 21.2 | 5.9 | 7.8 | 6.6 | |
| 12 | 40 | 0.006 | 10 | $3.3\times10^{6}$ | 35.8 | 1.6 | 21.9 | 6.0 | 7.8 | 6.7 | |
| 13 | 55 | 0.006 | 10 | $3.3\times10^{6}$ | 46.3 | 1.75 | 22.2 | 6.0 | 7.9 | 6.2 | $F_{XUVH}/F_{XUV}$ =1/3 |
| 14 | 7.5 | 0.005 | 10 | $3.3\times10^{6}$ | 7.5 | 0.77 | 19.2 | 6.1 | 7.4 | 6.9 | |
| 15 | 20 | 0.005 | 10 | $3.3\times10^{6}$ | 16.2 | 1.1 | 22.4 | 6.3 | 7.7 | 6.6 | |
| 16 | 25 | 0.003 | 10 | $3.3\times10^{6}$ | 21.6 | 0.8 | 22 | 6.6 | 8.2 | 6.9 | |
| 17 | 55 | 0.003 | 10 | $3.3\times10^{6}$ | 40.4 | 1.05 | 22.4 | 7.1 | 8.5 | 7.1 | |
| 18 | 25 | 0.003 | 250 | $3.3\times10^{6}$ | 26.6 | 0.81 | 20.7 | 6.2 | 9.0 | 7.4 | |
| 19 | 7.5 | 0.005 | 1000 | $3.3\times10^{6}$ | 7.5 | 0.55 | 16.4 | 6.1 | 7.4 | 6.5 | |
| 20 | 20 | 0.005 | 2000 | $3.3\times10^{6}$ | 16.2 | 0.82 | 17.0 | 6.3 | 7.7 | 7.0 | |
| 21 | 25 | 0.003 | 2000 | $3.3\times10^{6}$ | 21.6 | 0.69 | 16.9 | 6.6 | 8.2 | 7.5 | |
| 22 | 55 | 0.003 | 4000 | $3.3\times10^{6}$ | 40.4 | 0.90 | 18.0 | 7.1 | 8.5 | 3.3 | |
| Observations | | | | | | 1.1±0.1 [1*]<br>0.7±0.1 [2*] | 22.5 [1*]<br>27.2 [2*] | 14.4±3.6 [3*] | 7.7±2.7 [3*] | 5.0±1.3 [3*]<br>2.9±1.4 [4*] | |

Table 1. The list of simulation scenarios, with the corresponding modeling parameters, and calculated absorption features. Columns from left to right: number of the model run; assumed value of the integrated stellar XUV flux in the range of wavelengths 10Å <λ<910 Å in erg $cm^{-2}$ $s^{-1}$ at 1 a.u. (the hard energy part λ<510 Å of the flux is always taken as $F_{XUVH}$ =2/3·$F_{XUV}$ except the run N13); helium abundance in the base atmosphere; mass loss rate of the star which specifies the intensity of SW; cut-off density value which parametrizes the reduction of the Lyα cooling; calculated mass loss rate of the planet; calculated absorption maximum and half-maximum width of the simulated HeI($2^3$S) absorption profile at the mid-transit; simulated mid-transit Lyα absorption averaged over the blue wing [−230; −140] km/s and over the red wing [50; 230] km/s of the line; simulated total mid-transit Lyα absorption averaged over the whole line [-400; 400] km/s, with the exception of [-50; 50] km/s interval. The last column shows other modeling parameters varied additionally and described in the text. We chose specific absorption line averaging intervals to enable the comparison between the simulations with observations Bottom row indicates the measured absorption values provided in
($1^*$) Salz et al. (2018);
($2^*$) Guilluy et al. (2020);
($3^*$) *Des Etangs et al. (2012)* (the measurements in Sep. 2011, supported also with 2007/2008 measurements);
($4^*$) *Des Etangs et al. (2010)* (the measurements in Aug. 2010).

Figure 5 shows the excess absorption profiles at the position around 10830 Å caused by the HeI($2^3$S) triplet, simulated for different values of the stellar XUV flux. The considered XUV flux varies from about an average Solar value up to an order of magnitude higher than that. In the model run N13 we took the same features as in Lampón et al. (2021): XUV flux of $F_{XUV}$ =55 erg $cm^{-2}$/s at 1 a.u., with the hard flux at λ<510 Å three times less than the total (i.e., $F_{XUVH}/F_{XUV}$ =1/3). The helium abundance in the model runs presented in Figure 5 is fixed at He/H=0.006. One can see that the simulated absorption depth and, in lesser degree, the width of the absorption profile, increase with the increasing XUV flux. This is because of more efficient ionization of helium and faster expansion of the planetary atmosphere. Therefore, the factor of 1.5 difference between the measurements by Salz et al. (2018) and Guilluy et al. (2020) can be explained by an expected variability of the stellar XUV emission. In the right panel of Figure 5 we plot the results of two modeling scenarios aimed to fit these two different measurements. The first scenario assumes a very high XUV flux needed to reproduce the absorption of about 1% at the HeI($2^3$S) line center, as measured in Salz et al. (2018), and a relatively low helium abundance He/H=0.003. The second scenario takes a lower XUV flux, to fit absorption of about 0.7% at the line center, measured in Guilluy et al. (2020) and a bit higher helium abundance of He/H=0.005.

Altogether, the model run N16 provides relatively good correspondence with the measurements of Guilluy et al. (2020). Its reduced $\chi^2$ statistics value, calculated in the interval [-40; 20] km/s is 0.8, and it is the minimal value among all considered modeling runs. In calculating the $\chi^2$ we used for the degrees of freedom the number of observational data points minus the number of modeling parameters ($F_{XUV}$ and He/H).

Regarding the observations reported in Salz et al. (2018), the model does not reproduce the depth of the third line in the triplet, which appears to be larger than the stat-weight value of 1/8 relative to the sum of the first and second lines. Lampón et al. (2021) argued in that respect that the increasing atmosphere density (up to $10^{18}$ $cm^{-3}$) at the level close to the optical radius of the planet results in a stronger absorption at the optically thick region and, consequently, the smaller ratio of the 10832/10833 Å lines. However, from Figures 3 and 4 one can see that according to our modeling, in the dense region of the atmosphere of HD189733b at the altitudes $r<1.2R_p$, the HeI($2^3S$) atoms are much more efficiently depopulated rather than produced. It is worth to note that to include the optically thick region, we adopt in our model sufficiently high density of $10^{17}$ $cm^{-3}$ at the level of $r=R_p$.

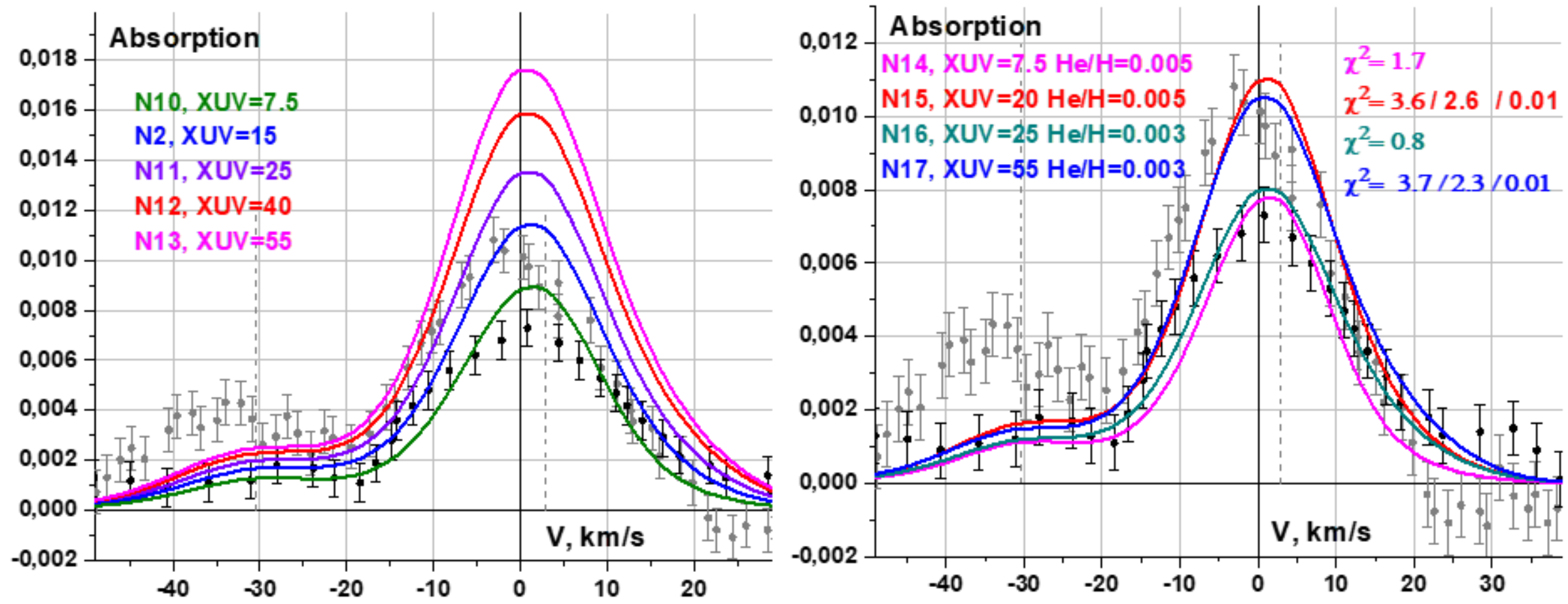

Figure 5. Left panel: HeI($2^3S$) triplet absorption profiles, simulated in the model runs N2, N10-13 with different XUV fluxes (specified in legend) and fixed helium abundance He/H=0.006, under conditions of a weak SW. Right panel: Reproducing of the two different measurements of HeI($2^3S$) triplet absorption (Salz et al. 2018, Guilluy et al. 2020) with the model runs N14-17, using two values of helium abundance and best fit XUV fluxes. The corresponding values of reduced $\chi^2$ calculated in the spectral interval of [-40; 20] km/s are provided in the legend. For the model runs N15 and N17 the provided second value of $\chi^2$ is calculated in the interval [-20; 20] km/s, whereas the third value was obtained in the interval [-40; 20] with the additional blue shift of the whole absorption profile by 2.5 km/s.

Another feature, which our model does not reproduce, is a net blue shift of about 2.5 km/s in the absorption data set in Salz et al. (2018), though such shift is absent in the data from Guilluy et al. (2020). While the escaping planetary material flow has a complex spiraling structure and asymmetry, it does not result in any significant shift of the line core, because the absorption comes from the region where velocity of the flow is still rather small. With several dedicated model runs, we checked if there might be any reasonable conditions, at which the radiation pressure acting on HeI($2^3S$) atoms can accelerate them up to velocities necessary for the observed blue shift. However, the stellar flux at 10830 Å appears to be well constrained by the SED computed for the given stellar parameters, and it appears too low to produce such acceleration during the average life-time of metastable atom.

An alternative way to explain the net blue shift in the HeI($2^3S$) absorption profile is related to the zonal flows from the dayside to the night side generated by the so-called hot spot in the atmosphere (Showman et al. 2015). Precise modeling of such flows requires a combination of our model of the escaping upper atmosphere of the planet with a GCM of its lower atmosphere. This challenging task is a subject of future work. At the moment, we can only try an empirical shift of the whole simulated HeI($2^3S$) absorption profile (in the model runs N15, N17), which results in the remarkably good correspondence with observations, characterized by sufficiently low $\chi^2$ =0.01.

To demonstrate the HeI($2^3S$) absorption in more details, Figure 6 shows the distribution over the stellar disk, as seen by the observer, calculated for three Doppler velocity intervals: the blue wing [-20; -7] km/s; the red wing [7; 20] km/s; the line core [-7; 7] km/s. One can see that the line core excess absorption is rather symmetric with a maximum value of about 15% outside the planet. Its appearance is restricted to the area within ~3$R_p$. The absorption in blue and red wings predominantly takes place behind and ahead the planet, respectively. This is related with the clock-wise twist of the escaping stream of upper atmospheric material under the action of Coriolis force. Therefore, the calculation of spectral absorption over the whole line, based on the predictions of simplified 1D models leads to the physically erroneous estimates and conclusions.

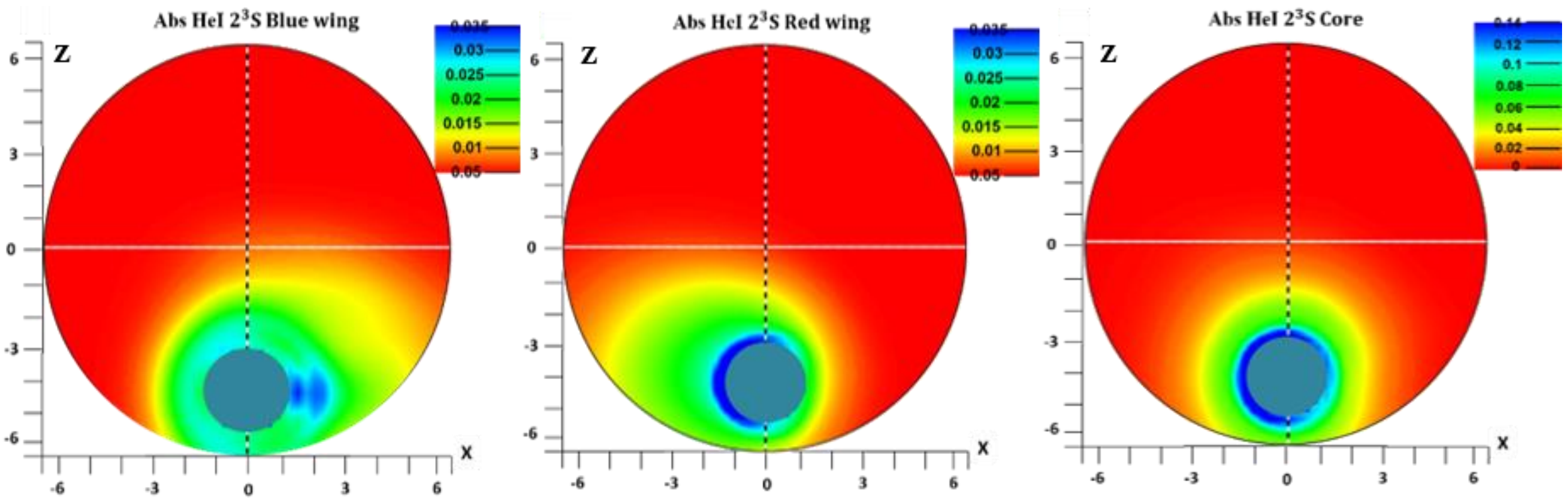

Figure 6. Distribution of the simulated in the model run N11 line-of-sight absorption of HeI($2^3S$) over the stellar disk (large red circle), calculated, under conditions of a weak SW for three velocity intervals (from left to right): [-20; -7] km/s; [7; 20] km/s; [-7; 7] km/s. The planet is indicated with a cyan circle.

Altogether, based on the performed modeling results, presented above, and comparison with observations, we conclude that in the considered case of HD189733b, the stellar XUV flux is constrained to the range of 7.5-55 erg $cm^{-2}$ /s at 1 a.u., whereas helium abundance should be He/H=(3-5)$\cdot 10^{-3}$. We explain the difference between peak absorption values measured in two different observation campaigns by possible variations of XUV flux, related with the activity of stars like HD189733.

Despite of the lower atmosphere of HD189733b dominantly consisting of the molecular hydrogen, it is rapidly dissociated at heights exposed to the stellar VUV and XUV fluxes. In the presence of even small sub-solar amounts of oxygen this dissociation proceeds much faster, facilitated by the molecular reactions with O and $H_2O$ (García Muñoz 2007), which are not included in our model. This is why we start our simulations already with the atomic hydrogen atmosphere imposed at the inner boundary base. However, the trial run with assumed molecular hydrogen base atmosphere, resulted in an order of magnitude smaller HeI($2^3S$) absorption and its significantly narrower profile. This is a consequence of the twice reduced in this case scale-height and more compact upper atmosphere. Therefore, our modeling results and their comparison with the measurements of HeI($2^3S$) absorption confirm the absence of the molecular hydrogen in the upper atmosphere of HD189733b, which has to be dissociated due to the complex molecular chemistry.

### 3.3 Strong stellar wind and the role of ENAs in Lyα absorptions

Now let us discuss the Lyα absorption. In that respect, our model produces the results similar to those of Odert et al. (2020). The absorption is generated by natural line broadening at the level of 5-7 % relatively close to the planet (<$3R_p$) in the region populated with the sufficiently high amount of the neutral atomic hydrogen. In the course of modeling we checked that the Lyα radiation pressure, acting on H atoms, does not accelerate them significantly up to the energies affecting the absorption profile. This is consistent with our previous studies dedicated to HD209458b (Shaikhislamov et al. 2020, Khodachenko et al. 2017) and GJ436b (Khodachenko et al. 2019). Regarding the HD189733b, an important question is whether the ENAs generated by charge exchange of planetary atoms with the SW protons can significantly add to the absorption at high blue-shifted velocities of the Lyα line, similar to that seen in the simulations of HD209458b, as well as in the observations and modeling of GJ436b (Lavie et al. 2017, Khodachenko et al. 2019) and GJ3470b (Bourrier et al. 2018, Shaikhislamov et al. 2021).

Figure 7 shows several Lyα absorption profiles simulated for different SW densities and stellar XUV fluxes, taking the helium abundances, which correspond to best fits of the HeI($2^3$S) absorption achieved in the model runs N14-N17. Under the conditions of a weak SW assumed in the model run N16, the simulated Lyα absorption profile appears to be very much symmetric. However, increasing intensity of the SW in the model runs N18-N22 results in a gradual increase of the absorption in the blue wing of the profile with respect to the red wing. This difference is related with the different character of interaction of the escaping planetary atmosphere and SW and the corresponding structure of material flows and the distribution of the absorbing atomic hydrogen. In particular, already in the model run N18 with the parameters of a moderate SW similar to those in an average solar wind, a well pronounced bowshock located upstream at ~$7R_p$ is formed, resulting in the production of ENAs and significant (up to 1.5 times) increase of the absorption at the range of Doppler shifted velocities around –75 km/s. However, in the range of higher velocities, [–230; –140] km/s, reported in Lecavelier Des Etangs et al. (2012), the input of ENAs is still negligible. Under the conditions of more intense SWs with higher densities and velocities, which might take place, e.g., in the CMEs, the bowshock region approaches extremely close to the planet, down to ~$3R_p$. This results in a qualitative change of the Lyα absorption character. First, the absorption depth at the line center drops up to several times, since the region of strong atmospheric absorption by the planetary hydrogen atoms shrinks down to an area with size of ~$3R_p$ around the planet, whereas under a weak SW conditions the full size of this region was ~$5R_p$. At the same time, significant (≈15%) Lyα absorption at the blue range of Doppler shifted velocities around –100 km/s appears. Only about half of this absorption is due to ENAs. The rest part comes from the planetary hydrogen atoms, which are not involved in the charge-exchange reaction with the SW protons, but are collisionally energized in the shock region where the density of SW protons is ~$10^6$ $cm^{-3}$, and the planetary atoms are picked-up by the SW flow. Similar effect was also found in Khodachenko et al. (2017) while studying with a 2D HD model the interaction of hot Jupiter planetary wind with different stellar winds.

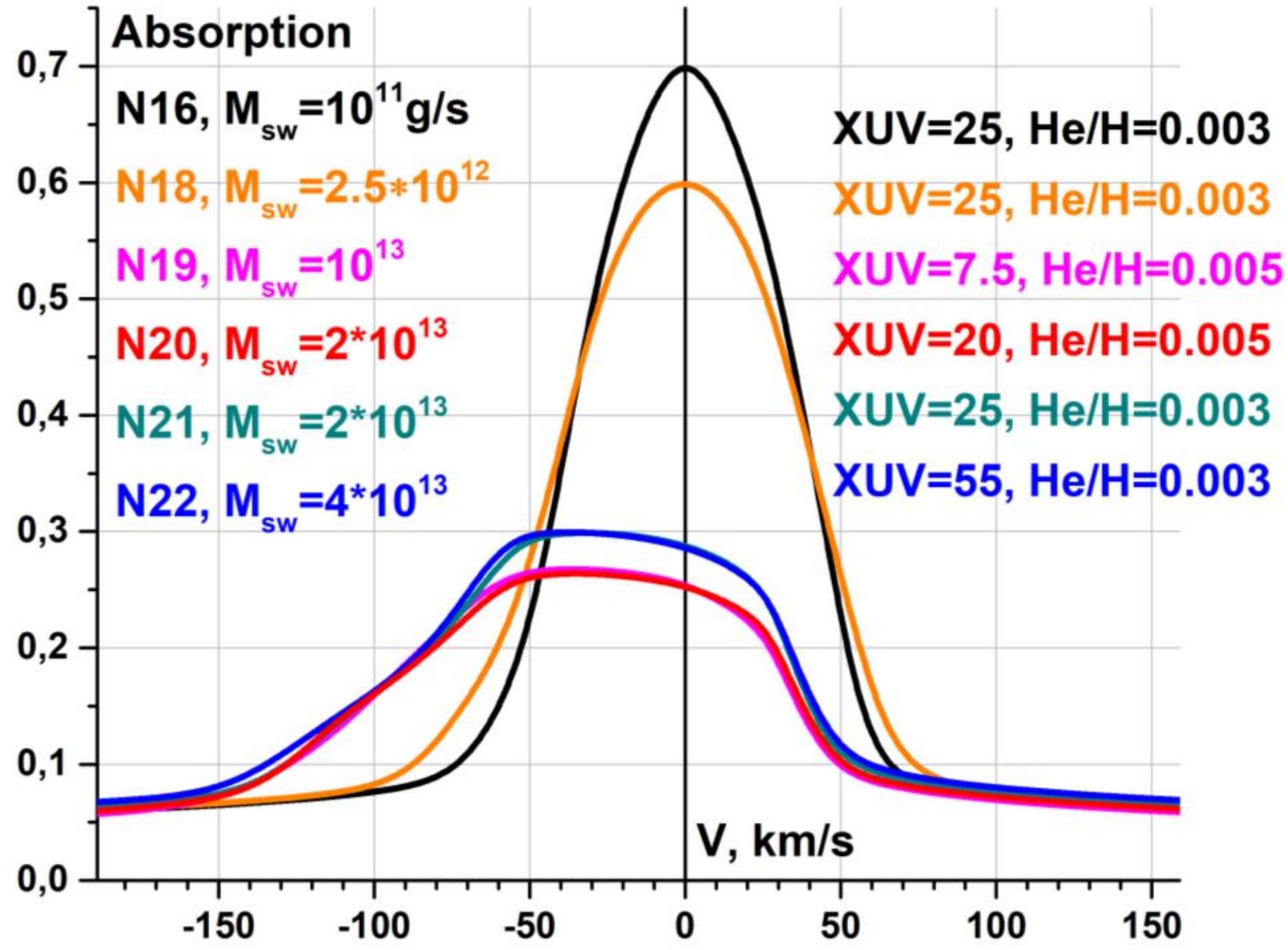

Figure 7. Lyα line absorption profiles simulated under conditions of week (run N16), moderate (run N18) and strong (runs N19-22) SW.

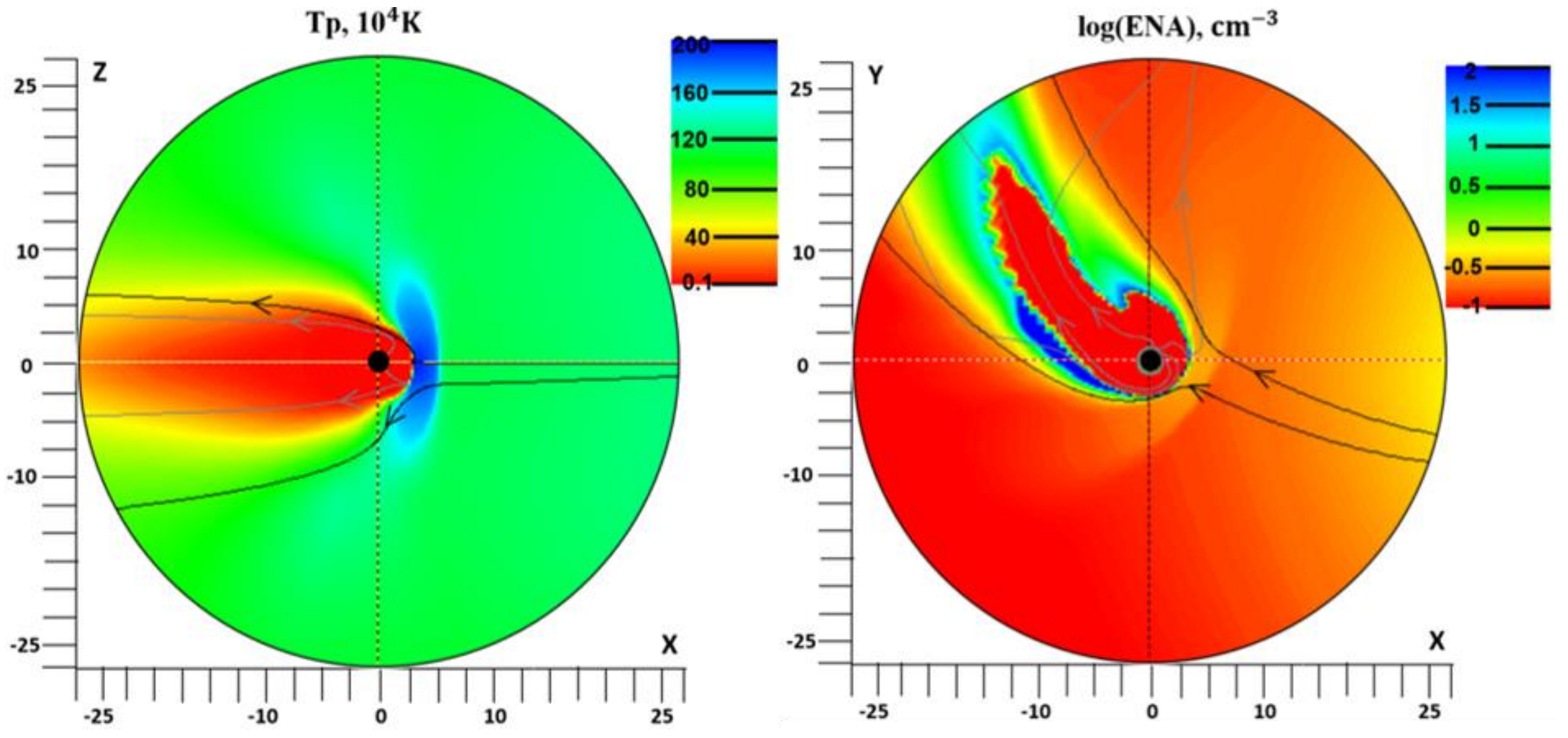

Figure 8. Distribution of the proton temperature (left panel, meridional plane) and ENA density (right panel, orbital plane) around HD189733b, simulated under the conditions of strong SW in the model run N21. (The stellar protons streamlines are shown as black lines, and the streamlines of planetary protons (in the left panel) and planetary atoms (in the right panel) are colored gray.

Figure 8 shows the proton temperature and ENA density distributions around the planet simulated in the model run N21 under the conditions of strong SW, which are similar to the case of CMEs considered in Odert et al. (2020). These plots clearly show the bowshock formed around the planet and details of the planetary and stellar material motion. According to Figure 7, the effect of ENAs is well pronounced in the simulated Lyα absorption, which was not the case in Odert et al. (2020). One of the reasons for this difference is that we model the interaction of the planetary and stellar material flows in full 3D, and the most ENAs are produced in the tail region, which is absent in the simulations of Odert et al. (2020).

To shed more light on the origin of the simulated and measured Lyα absorption in HD189733b, we present in Figure 9 distribution of this absorption across the stellar disk in different Doppler velocity intervals, and produced by different populations of particles. The absorption over the whole line in the range of [-150; 150] km/s, except of the geo-contaminated interval (±50 km/s) is more or less symmetric. It originates from a relatively close projected region around the planet within $\approx 3R_p$. However, it is worth to note that this projected region contains also an additional input from the hydrogen atoms in the tail. The input of tail particles becomes more evident at the blue wing of the Lyα line, especially in the part of absorption contributed by the ENAs generated by the charge-exchange reaction. These tail particles are clearly seen in Figure 8 (right panel).

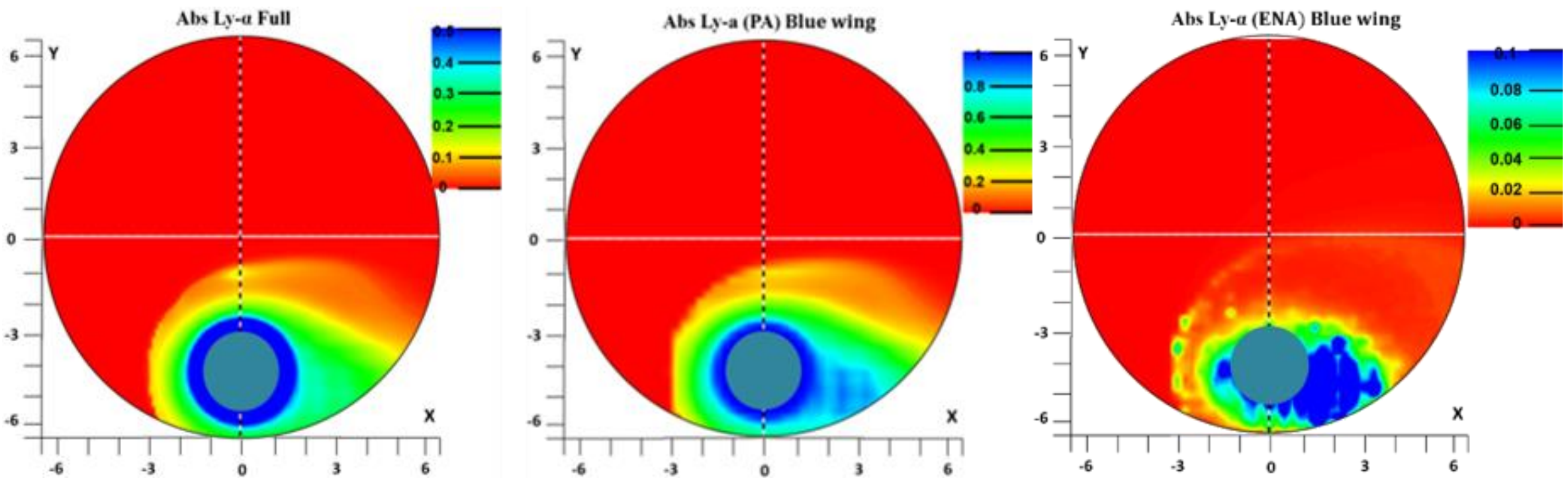

Figure 9. Distribution of the simulated in the model run N21 line-of-sight Lyα absorption by hydrogen atoms over the stellar disk (large red circle), calculated for different velocity intervals and particle populations. From left to right: 1) full absorption over [-150; 150] km/s, excluding the geo-contaminated ±50 km/s interval; 2) blue wing [-150; -50] km/s absorption produced by the collisionally energized planetary hydrogen atoms, picked-up by the SW; 3) the blue wing [-150; -50] km/s absorption produced by the ENAs generated via charge exchange of stellar protons with planetary atoms. The planet is indicated with a cyan circle.

Finally, in Figure 10 we present how the strong SW might affect the HeI($2^3$S) line absorption. One can see that while the moderate SW does not influence the absorption profile, the strong SW significantly reduces the absorption in the red wing, as well as the overall width and depth of the profile. This result leads us to the conclusion that at the time of the discussed in this paper measurements of the HeI($2^3$S) absorption, the SW could not be strong.

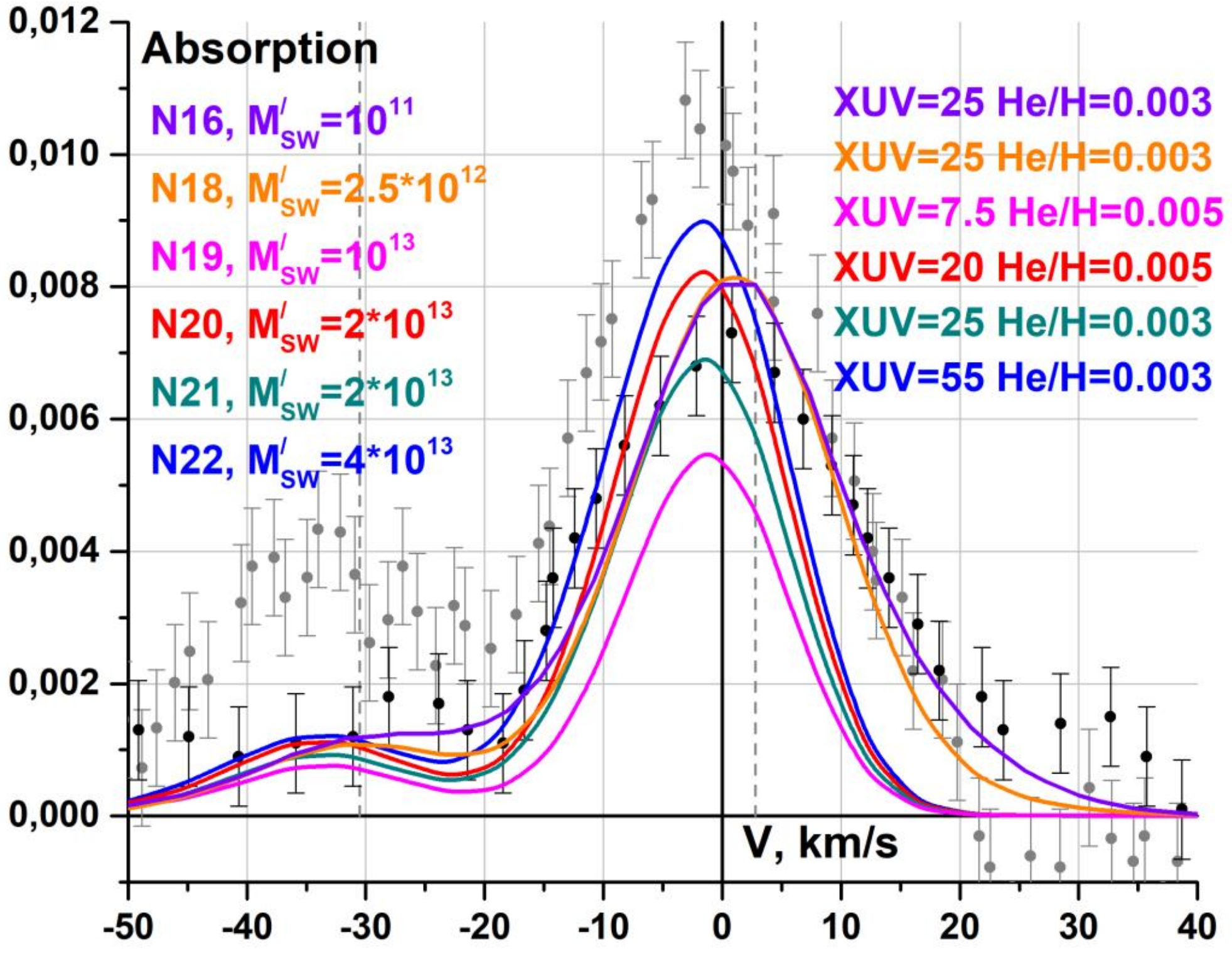

Figure 10. HeI($2^3$S) triplet absorption profiles, simulated under conditions of a moderate (run N18) and strong (runs N19-22) SW.

## 4. Conclusions

Despite a number of FUV spectral observations performed with HST, the existing stellar activity and variability prevent an accurate constraining of the parameters of the upper atmosphere of the hot Jupiter HD189733b. While the measurements of 5% depth in the whole Lyα line are interpreted within the frame of 1D HD models as the natural line broadening absorption by the planetary hydrogen in the expanding atmosphere confined in the region of a few planetary radii, the measurement of 14% depth at the high velocity >140 km/s blue wing of the Lyα line obtained in a single observational session remains enigmatic for hydrodynamic models. Bourrier et al. (2013) reports that the Monte Carlo model is able to fit this particular observation by the radiation pressure acceleration of the atomic hydrogen and charge exchange with SW protons. It is obtained at planetary mass losses $4 \cdot 10^8 – 4 \cdot 10^9$ g/s, which are significantly lower than those predicted by gasdynamic models at similar XUV fluxes of 10-30 erg cm$^{-2}$/s at 1 a.u. At the same time, according to Bourrier et al. (2013), the stellar wind at the planet orbit should have velocity of 200 km/s and rather low temperature of $3 \cdot 10^4$ K, while density can vary in a wide range around $10^5$ cm$^{-3}$. However, the 3D HD modeling in Odert et al. (2020), as well as that presented in this paper, do not confirm the generation of ENAs with the velocities in excess of 140 km/s by either of the mechanisms. At the same time, we found a significant increase of the Lyα absorption in the range [-50; -150] km/s, with an average absorption value up to 16% under the conditions of a strong SW, characterized by velocity of 240 km/s and density of >$10^5$ cm$^{-3}$, while the temperature, as dictated by our empirical heating model, should be around $10^6$ K. Note that the measured by HST the absorption of 13.2% at velocities of [-117; -156] km/s at the 3.11.2013 visit (Bourrier et al. 2020), agrees much better with our simulation. One of the physical features, which can significantly influence the interaction of planetary and stellar particles and interpretation of observations, is the planetary and stellar magnetic fields (Pillitteri et al. 2015, Khodachenko et al. 2021b). Investigation of this possibility remains a subject of future studies.

The measurements of absorption at the position of the metastable HeI($2^3$S) triplet line, performed with the ground-based telescopes, offered a new possibility to verify the existing models and physical scenarios and to constrain the parameters of the HD189733b system. With the assumed cooling of the planetary upper atmosphere by excitation of the hydrogen atoms, the simulated width of the HeI($2^3$S) absorption appeared significantly narrower than that in the observations. This is because the upper atmosphere of the HD189733b under such an assumed cooling is in a radiative balance rather than in an outflow regime. Salz et al. (2016) reported the same conclusion and approximately similar parameters of planetary wind. Within the frame of our model we could reproduce the width of the measured HeI($2^3$S) absorption profile by reducing the overall effect of Lyα cooling in the energy balance by about two times. This increases by about 2.5 times the part of energy going to the acceleration of the escaping atmosphere flow and increases the mass loss rate by an order of magnitude. Note that, at the same time, the absorption in Lyα line is only slightly affected by the cooling.

With the reduced Lyα cooling, our simulations indicates that the helium abundance could not be larger than He/H=0.01, otherwise the synthetic absorption profiles become too narrow to match the observations. The peak HeI($2^3$S) absorption is matched in the range of stellar XUV fluxes (7.5-55) erg $cm^{-2}$/s at 1 a.u., whereas the helium abundance is constrained within the range of (3-5)$\cdot 10^{-3}$. Our results are in better agreement with the reconstructed SED of Bourrier et al. 2020 at lower He/H ratio (run N16). We explain the factor of 1.5 difference between two available measurements of the HeI($2^3$S) absorption by 2-2.5 times natural fluctuations of the stellar XUV flux. Another constrain that our modeling enables to put, based on the comparison of the measured and simulated HeI($2^3$S) absorption profiles, is that the SW could not be very strong during the observations.

Remarkably, the best-fit result regarding the helium abundance, (5–8)$\cdot 10^{-3}$ in Lampón et al. (2021), obtained at the mass loss rate of $1.1\cdot 10^{11}$ g/s, is quite close to the predictions of our global 3D HD modeling. However, the approaches to derive these values are principally different. In fact, the mass loss rate and temperature of the escaping planetary flow are not the free parameters of our model. They are calculated in the simulations in dependence on more fundamental and independent physical parameters of the system, namely, the integral stellar XUV flux, which is a subject to temporal variability, and helium abundance, which is not known and has to be constrained. It turned out that the Lyα absorption, which is to a significant extent is produced by the planetary hydrogen due to the natural line broadening, is quite insensitive to XUV flux and the related with it atmospheric mass loss. Altogether, our simulations show that the Lyα absorption cannot be used to derive the planetary mass loss and the helium abundance, as it was done in Lampón et al. (2021).

In general, our simulations show that the global 3D fully self-consistent multi-component modeling produces the results consistent with observations at transits of HD189733b in Lyα and HeI($2^3$S) lines at physical parameters which are well within the expected ranges. The synthetic profiles simulated with our model for the HeI($2^3$S) absorption do not fit the observation data points as close as they do in Lampón et al. (2021). This is because our self-consistent modeling approach has a limited number of free parameters, specifically the XUV radiation flux with its energy partition $F_{XUXH}/F_{XUV}$, elements abundance, as well as SW parameters and efficiency of Lyα cooling. Therefore, the insignificant difference between the simulations and measurements in our case is more likely because of additional physical effects and processes not covered by the model, which nevertheless well reproduces the major observational phenomena. Also, we do not reproduce the net blue shift of the HeI($2^3$S) absorption profile by a few km/s, though it greatly improves the fit between the modeling and observations.

At the same time, we can add in our model one more free parameter, assuming that the planet is not completely tidally locked, but rotates, or there exists in the atmosphere a strong eastward zonal jet (Showman et al. 2015). We checked this possibility versus the observations and obtained that the rotational azimuthal velocities at the inner boundary of the simulation domain, i.e., at the base of the modeled upper atmosphere, might influence the width of the absorption profile, given they are of sufficiently large values of more than 10 km/s. However, detailed conclusion on the role of the zonal winds is a subject for a special study, because the parameters of an atmospheric winds cannot be formally inserted in a self-consistent model to produce the desired effect. All the assumptions and prescribed features should have firm physical grounds, or be provided as an outcome of other dedicated and integrated models.

**Acknowledgements:**

This work was performed under the grant № 18-12-00080 of the Russian Science Foundation. Analysis of Lyα cooling effect was accomplished in the frame of the of grant № 21-72-00129 of the Russian Science Foundation. MSR and AGB are thankful to the project "Study of stars with exoplanets" under a grant from the Government of the Russian Federation for scientific research conducted under the guidance of leading scientists (agreement N 075-15-2019-1875) for supporting investigation of Lyα absorption properties and additional rotation affected on HeI($2^3$S) absorption properties. MLK acknowledges the projects I2939-N27 and S11606-N16 of the Austrian Science Fund (FWF). HL acknowledges the projects P25256-N27 and S11607-N16 of the Austrian Science Foundation (FWF). IFS is thankful to the Programme “Astrophysical Origins: Pathways from Star Formation to Habitable Planets” (July 2019, ESI, Wien, Austria) for productive discussions. IBM and AGB acknowledge the support of RFBR grant № 20-02-00520. Parallel computing simulations, key for this study, have been performed at Computation Center of Novosibirsk State University, SB RAS Siberian Supercomputer Center and Joint Supercomputer Center of RAS.

**References**

Allart, R., Bourrier, V., Lovis, C., Ehrenreich, D., Spake, J. J., Wyttenbach, A., ... & des Etangs, A. L. «Spectrally resolved helium absorption from the extended atmosphere of a warm Neptune-mass exoplanet». Science, 362(6421), 2018: 1384-1387.

Allart, R., Bourrier, V., Lovis, C., Ehrenreich, D., Aceituno, J., Guijarro, A., ... & Wyttenbach, A. «High-resolution confirmation of an extended helium atmosphere around WASP-107b». Astronomy & Astrophysics, 623, 2019: A58.

Alonso-Floriano, F. J., Snellen, I. A. G., Czesla, S., Bauer, F. F., Salz, M., Lampón, M., ... & Osorio, M. Z. «He I λ 10 830 Å in the transmission spectrum of HD209458 b». Astronomy & Astrophysics, 629, 2019: A110.

Aver, E., Olive, K. A., & Skillman, E. D. «The effects of He I λ10830 on helium abundance determinations». Journal of Cosmology and Astroparticle Physics, 2015(07), 2015: 011.

Ben-Jaffel, L. «Exoplanet HD 209458b: Inflated hydrogen atmosphere but no sign of evaporation». The Astrophysical Journal Letters, 671(1), 2007: L61.

Bouchy, F., Udry, S., Mayor, M., Moutou, C., Pont, F., Iribarne, N., ... & Zucker, S. «ELODIE metallicity-biased search for transiting hot jupiters-II. a very hot jupiter transiting the bright k star HD 189733». Astronomy & Astrophysics, 444(1), 2005: L15.

Bourrier, V., & des Etangs, A. L. «3D model of hydrogen atmospheric escape from HD 209458b and HD 189733b: radiative blow-out and stellar wind interactions». Astronomy & Astrophysics, 557, 2013a: A124.

Bourrier, V., des Etangs, A. L., Dupuy, H., Ehrenreich, D., Vidal-Madjar, A., Hébrard, G., ... & Wheatley, P. J. «Atmospheric escape from HD 189733b observed in H i Lyman-α: detailed analysis of HST/STIS September 2011 observations». Astronomy & Astrophysics, 551, 2013b: A63.

Bourrier, V., Des Etangs, A. L., Ehrenreich, D., Sanz-Forcada, J., Allart, R., Ballester, G. E., ... & Wakeford, H. R. «Hubble PanCET: an extended upper atmosphere of neutral hydrogen around the warm Neptune GJ 3470b». Astronomy & Astrophysics, 620, 2018: A147.

Bourrier, V., Wheatley, P. J., Lecavelier des Etangs, A., King, G., Louden, T., Ehrenreich, D., ... & Vidotto, A. A. (2020). MOVES III. Simultaneous X-ray and ultraviolet observations unveiling the variable environment of the hot Jupiter HD 189733b. *Monthly Notices of the Royal Astronomical Society*, *493*(1), 559-579.

Chadney, J. M., Koskinen, T. T., Galand, M., Unruh, Y. C., & Sanz-Forcada, J. (2017). Effect of stellar flares on the upper atmospheres of HD 189733b and HD 209458b. *Astronomy & Astrophysics*, *608*, A75.

Charbonneau, D., Brown, T. M., Latham, D. W., & Mayor, M. «Detection of planetary transits across a sun-like star». The Astrophysical Journal Letters, 529(1), 2000: L45.

Christie, D., Arras, P., & Li, Z. Y. «Hα absorption in transiting exoplanet atmospheres». The Astrophysical Journal, 772(2), 2013: 144.

Cubillos, P. E., Fossati, L., Koskinen, T., Young, M. E., Salz, M., France, K., ... & Haswell, C. A. «Near-ultraviolet transmission spectroscopy of HD 209458b: evidence of ionized iron beyond the planetary Roche lobe». The Astronomical Journal, 2020.

Des Etangs, A. L., Ehrenreich, D., Vidal-Madjar, A., Ballester, G. E., Désert, J. M., Ferlet, R., ... & Udry, S. «Evaporation of the planet HD 189733b observed in H i Lyman-α». Astronomy & Astrophysics, 514, 2010: A72.

Des Etangs, A. L., Bourrier, V., Wheatley, P. J., Dupuy, H., Ehrenreich, D., Vidal-Madjar, A., ... & Sing, D. K. «Temporal variations in the evaporating atmosphere of the exoplanet HD 189733b». Astronomy & Astrophysics, 543, 2012: L4.

Dwivedi, N. K., Khodachenko, M. L., Shaikhislamov, I. F., Fossati, L., Lammer, H., Sasunov, Y., ... & Güdel, M. (2019). Modeling atmospheric escape and Mg ii near-ultraviolet absorption of the highly irradiated hot Jupiter WASP-12b. *Monthly Notices of the Royal Astronomical Society*, *487*(3), 4208-4220.

Guilluy, G., Andretta, V., Borsa, F., Giacobbe, P., Sozzetti, A., Covino, E., ... & Reiners, A. «The GAPS programme at TNG-XXII. The GIARPS view of the extended helium atmosphere of HD 189733 b accounting for stellar activity». Astronomy & Astrophysics, 639, 2020: A49.

Guo, J. H. «Escaping particle fluxes in the atmospheres of close-in exoplanets. I. Model of hydrogen». The Astrophysical Journal, 733(2), 2011: 98

Huang, C., Arras, P., Christie, D., & Li, Z. Y. «A model of the Hα and Na transmission spectrum of HD 189733b». The Astrophysical Journal, 851(2), 2017: 150.

Khodachenko, M. L., Shaikhislamov, I. F., Lammer, H., Kislyakova, K. G., Fossati, L., Johnstone, C. P., ... & Posukh, V. G. (2017). Lyα absorption at transits of HD 209458b: a comparative study of various mechanisms under different conditions. *The Astrophysical Journal*, *847*(2), 126.

Khodachenko, M. L., Shaikhislamov, I. F., Lammer, H., Berezutsky, A. G., Miroshnichenko, I. B., Rumenskikh, M. S., ... & Dwivedi, N. K. «Global 3D Hydrodynamic Modeling of In-transit Lyα Absorption of GJ 436b». The Astrophysical Journal, 885(1), 2019: 67.

Khodachenko, M. L., Shaikhislamov, I. F., Fossati, L., Lammer, H., Rumenskikh, M. S., Berezutsky, A. G., ... & Efimof, M. A. «Simulation of 10 830 Å absorption with a 3D hydrodynamic model reveals the solar He abundance in upper atmosphere of WASP-107b». Monthly Notices of the Royal Astronomical Society: Letters, 503(1), 2021a: L23-L27.

Khodachenko M.L., Shaikhislamov, I.F., Lammer, H., Miroshnichenko, I.B., Rumenskikh, M.S., Berezutsky, A.G., Fossati, L., The impact of intrinsic magnetic field on the absorption signatures of elements probing the upper atmosphere of HD209458b, MNRAS, 2021b, 507(3), 3626–3637 (DOI: 10.1093/mnras/stab2366)

Koskinen, T. T., Aylward, A. D., Smith, C. G. A., & Miller, S. «A thermospheric circulation model for extrasolar giant planets». The Astrophysical Journal, 661(1), 2007: 515.

Lammer, H., Selsis, F., Ribas, I., Guinan, E. F., Bauer, S. J., & Weiss, W. W. «Atmospheric loss of exoplanets resulting from stellar X-ray and extreme-ultraviolet heating». The Astrophysical Journal Letters, 598(2), 2003: L12.

Lampón, M., López-Puertas, M., Sanz-Forcada, J., Sánchez-López, A., Molaverdikhani, K., Czesla, S., ... & Ribas, I. «Modeling the He I triplet absorption at 10 830 Å in the atmospheres of HD 189733 b and GJ 3470 b». Astronomy & Astrophysics, 647, 2021: A129.

Leighly, K. M., Dietrich, M., & Barber, S. «The discovery of the first He iλ10830 broad absorption line quasar». The Astrophysical Journal, 728(2), 2011: 94.

Lin, H., Penn, M. J., & Kuhn, J. R. «He I 10830 Å line polarimetry: A new tool to probe the filament magnetic fields». The Astrophysical Journal, 493(2), 1998: 978.

Linsky, J. L., Yang, H., France, K., Froning, C. S., Green, J. C., Stocke, J. T., & Osterman, S. N. «Observations of mass loss from the transiting exoplanet HD 209458b». The Astrophysical Journal, 717(2), 2010: 1291.

Mansfield, M., Bean, J. L., Oklopčić, A., Kreidberg, L., Désert, J. M., Kempton, E. M. R., ... & Bourrier, V. «Detection of Helium in the Atmosphere of the Exo-Neptune HAT-P-11b». The Astrophysical Journal Letters, 868(2), 2018: L34.

Menager, H., Barthélemy, M., Koskinen, T., Lilensten, J., Ehrenreich, D., & Parkinson, C. D. «Calculation of the H Lyman α emission of the hot Jupiters HD 209458b and HD 189733b». Icarus, 226(2), 2013: 1709-1718.

Muñoz, A. G. «Physical and chemical aeronomy of HD 209458b». Planetary and Space Science, 55(10), 2007: 1426-1455.

Murray-Clay, R. A., Chiang, E. I., & Murray, N. «Atmospheric escape from hot Jupiters». The Astrophysical Journal, 693(1), 2009: 23.

Ninan, J. P., Stefansson, G., Mahadevan, S., Bender, C., Robertson, P., Ramsey, L., ... & Schwab, C. «Evidence for He I 10830 Å Absorption during the Transit of a Warm Neptune around the M-dwarf GJ 3470 with the Habitable-zone Planet Finder». The Astrophysical Journal, 894(2), 2020: 97.

Nortmann, L., Pallé, E., Salz, M., Sanz-Forcada, J., Nagel, E., Alonso-Floriano, F. J., ... & Osorio, M. R. Z. «Ground-based detection of an extended helium atmosphere in the Saturn-mass exoplanet WASP-69b». Science, 362(6421), 2018: 1388-1391.

Odert, P., Erkaev, N. V., Kislyakova, K. G., Lammer, H., Mezentsev, A. V., Ivanov, V. A., ... & Holmstroem, M. «Modeling the Lyα transit absorption of the hot Jupiter HD 189733b». Astronomy & Astrophysics, 638, 2020: A49.

Oklopčić, A., & Hirata, C. M. «A new window into escaping exoplanet atmospheres: 10830 Å line of helium». The Astrophysical Journal Letters, 855(1), 2018: L11.

Palle, E., Nortmann, L., Casasayas-Barris, N., Lampón, M., López-Puertas, M., Caballero, J. A., ... & Zechmeister, M. «A He I upper atmosphere around the warm Neptune GJ 3470 b». Astronomy & Astrophysics, 638, 2020: A61.

Pillitteri, I., Maggio, A., Micela, G., Sciortino, S., Wolk, S. J., & Matsakos, T. «FUV variability of HD 189733. Is the star accreting material from its hot Jupiter?» The Astrophysical Journal, 805(1), 2015: 52.

Salz, M., Czesla, S., Schneider, P. C., & Schmitt, J. H. M. M. «Simulating the escaping atmospheres of hot gas planets in the solar neighborhood». Astronomy & Astrophysics, 586, 2016: A75.

Salz, M., Czesla, S., Schneider, P. C., Nagel, E., Schmitt, J. H. M. M., Nortmann, L., ... & Zechmeister, M. «Detection of He I λ10830 Å absorption on HD 189733 b with CARMENES high-resolution transmission spectroscopy». Astronomy & Astrophysics, 62, 2018: A97.

Seager, S., & Sasselov, D. D. «Theoretical transmission spectra during extrasolar giant planet transits». The Astrophysical Journal, 537(2), 2000: 916.

Shaikhislamov, I. F., Khodachenko, M. L., Sasunov, Y. L., Lammer, H., Kislyakova, K. G., & Erkaev, N. V. «Atmosphere expansion and mass loss of close-orbit giant exoplanets heated by stellar XUV. I. Modeling of hydrodynamic escape of upper atmospheric material». The Astrophysical Journal, 795(2), 2014: 132.

Shaikhislamov, I. F., Khodachenko, M. L., Lammer, H., Kislyakova, K. G., Fossati, L., Johnstone, C. P., ... & Posukh, V. G. «Two regimes of interaction of a Hot Jupiter's escaping atmosphere with the stellar wind and generation of energized atomic hydrogen corona». The Astrophysical Journal, 832(2), 2016: 173.

Shaikhislamov, I. F., Khodachenko, M. L., Lammer, H., Berezutsky, A. G., Miroshnichenko, I. B., & Rumenskikh, M. S. «3D Aeronomy modeling of close-in exoplanets». Monthly Notices of the Royal Astronomical Society, 481(4), 2018a: 5315-5323.

Shaikhislamov, I. F., Khodachenko, M. L., Lammer, H., Fossati, L., Dwivedi, N., Güdel, M., ... & Ivanov, V. A. (2018b). Modeling of Absorption by Heavy Minor Species for the Hot Jupiter HD 209458b. *The Astrophysical Journal*, *866*(1), 47.

Shaikhislamov, I. F., Khodachenko, M. L., Lammer, H., Berezutsky, A. G., Miroshnichenko, I. B., & Rumenskikh, M. S. «Three-dimensional modeling of absorption by various species for hot Jupiter HD 209458b». Monthly Notices of the Royal Astronomical Society, 491(3), 2020a: 3435-3447.

Shaikhislamov, I. F., Fossati, L., Khodachenko, M. L., Lammer, H., Muñoz, A. G., Youngblood, A., ... & Rumenskikh, M. S. «Three-dimensional hydrodynamic simulations of the upper atmosphere of π Men c: Comparison with Lyα transit observations». Astronomy & Astrophysics, 639, 2020b: A109.

Shaikhislamov, I. F., Khodachenko, M. L., Lammer, H., Berezutsky, A. G., Miroshnichenko, I. B., & Rumenskikh, M. S. «Global 3D hydrodynamic modeling of absorption in Lyα and He 10830 Å lines at transits of GJ3470b». Monthly Notices of the Royal Astronomical Society, 500(1), 2021: 1404-1413.

Showman, A. P., & Polvani, L. M. «Equatorial superrotation on tidally locked exoplanets». The Astrophysical Journal, 738(1), 2011: 71.

Showman, A. P., Lewis, N. K., & Fortney, J. J. «Three-dimensional atmospheric circulation of warm and hot Jupiters: Effects of orbital distance, rotation period, and nonsynchronous rotation». The Astrophysical Journal, 801(2), 2015: 95.

Shulyak, D., Tsymbal, V., Ryabchikova, T., Stütz, C., & Weiss, W. W. «Line-by-line opacity stellar model atmospheres». Astronomy & Astrophysics, 428(3), 2004: 993-1000.

Vidal-Madjar, A., Des Etangs, A. L., Désert, J. M., Ballester, G. E., Ferlet, R., Hébrard, G., & Mayor, M. «An extended upper atmosphere around the extrasolar planet HD209458b». Nature, 422(6928), 2003: 143-146

Vidal-Madjar, A., Désert, J. M., Des Etangs, A. L., Hébrard, G., Ballester, G. E., Ehrenreich, D., ... & Parkinson, C. D. «Detection of oxygen and carbon in the hydrodynamically escaping atmosphere of the extrasolar planet HD 209458b». The Astrophysical Journal, 2004.

Vidal-Madjar, A., Des Etangs, A. L., Désert, J. M., Ballester, G. E., Ferlet, R., Hébrard, G., & Mayor, M. «Exoplanet HD 209458b (Osiris*): Evaporation Strengthened». The Astrophysical Journal Letters, 676(1), 2008: L57.

Weber, C., Erkaev, N. V., Ivanov, V. A., Odert, P., Grießmeier, J. M., Fossati, L., ... & Rucker, H. O. «Supermassive hot Jupiters provide more favourable conditions for the generation of radio emission via the cyclotron maser instability–a case study based on Tau Bootis b». Monthly Notices of the Royal Astronomical Society, 480(3), 2018: 3680-3688.

Yelle, R. V. «Aeronomy of extra-solar giant planets at small orbital distances». Icarus, 170(1), 2004: 167-179.